\documentclass{emulateapj}
\usepackage{longtable}
\usepackage{rotating}
\usepackage{natbib,times}
\usepackage{threeparttable}
\usepackage{amsmath}

\newcommand{\mc}{\multicolumn}

\shorttitle{Scaling relations for radio halos and relics of clusters} 
\shortauthors{Yuan, Han, \& Wen}
\begin{document}

\title{The scaling relations and the fundamental plane for radio halos
  and relics of galaxy clusters}

\author{Z. S. Yuan$^{1,2}$, J. L. Han$^1$, \and Z. L. Wen$^1$}
\affil{$^1$National Astronomical Observatories, 
                Chinese Academy of Sciences, 
                20A Datun Road, Chaoyang District, Beijing 100012, China; 
                hjl@nao.cas.cn\\
       $^2$School of Physics,
                University of Chinese Academy of Sciences, 
                Beijing 100049, China}

\begin{abstract}
Diffuse radio emission in galaxy clusters is known to be related to
cluster mass and cluster dynamical state. We collect the observed
fluxes of radio halos, relics, and mini-halos for a sample of galaxy
clusters from the literature, and calculate their radio powers. We
then obtain the values of cluster mass or mass proxies from previous
observations, and also obtain the various dynamical parameters of
these galaxy clusters from optical and X-ray data. The radio powers of
relics, halos, and mini-halos are correlated with the cluster masses
or mass proxies, as found by previous authors, with the correlations
concerning giant radio halos being, in general, the strongest ones. We
found that the inclusion of dynamical parameters as the third
dimension can significantly reduce the data scatter for the scaling
relations, especially for radio halos. We therefore conclude that the
substructures in X-ray images of galaxy clusters and the irregular
distributions of optical brightness of member galaxies can be used to
quantitatively characterize the shock waves and turbulence in the
intracluster medium responsible for re-accelerating particles to
generate the observed diffuse radio emission. The power of radio halos
and relics is correlated with cluster mass proxies and dynamical
parameters in the form of a fundamental plane.
\end{abstract}

\keywords{galaxies: clusters: general --- 
          galaxies: clusters: intracluster medium}

\section{Introduction}

Clusters of galaxies are the largest gravitationally bound systems in
the universe, formed at knots of cosmic webs in the universe. Diffuse
radio emission has been detected from about 100 galaxy clusters. Based
on their morphology, location, and size, the diffuse radio sources are
classified as radio halos, radio relics, or mini-halos (see
\citealt{FGG+12} for an observational review). Radio halos are located
at the cluster center and unpolarized ($<10\%$), and have a regular
morphology with a typical scale of about 1 Mpc. Radio relics usually
also have a similar size, but are located in peripheral regions of
galaxy clusters and often polarized ($\sim$20-30$\%$). Mini-halos are
detected in the central regions of clusters with no obvious
polarization, but have a smaller size ($\lesssim$500 kpc).  Radio
halos and relics are clearly related to cluster mergers
\citep[e.g.,][]{CEGB+10,CCB+15}, while mini-halos are detected in
cool-core galaxy clusters \citep[e.g.,][]{CEGB+10,VRB+10,KVG+15}.
Both mini-halos and giant radio halos are detected from clusters with
high X-ray luminosity \citep{KVG+15}. Observations of diffuse radio
emission of clusters open a new window to study the intracluster
medium, especially the particle accelerations and magnetic field
amplification of galaxy clusters with different dynamical states
(see \citealt{BJ14} for a review).

Strong correlations have been found between the radio power at 1.4
GHz, $P_{\rm 1.4~GHz}$, of radio halos and other physical cluster
parameters, namely the cluster X-ray luminosity, $L_{\rm X}$, and hot
gas temperature, $T_{\rm X}$
\citep[e.g.,][]{LHB+00,BVD+07,BCD+09,CEB+13}.  Mini-halos also follow
a similar relation between radio power and the cluster X-ray
luminosity \citep{CGB+08,KVG+13,KVG+15}. Correlations between $P_{\rm
  1.4~GHz}$ of radio relics and the cluster X-ray luminosity have also
been found \citep[e.g.,][]{FGG+12,DVB+14}.  However, radio halos are
detected only from only $20\%$ to $30\%$ of massive clusters with high
X-ray luminosity \citep[e.g.,][]{KVG+13,KVG+15}.
\citet{BVD+07,BCD+09} first discovered this radio bimodality that
galaxy clusters with radio halos follow the correlation between the
radio power and cluster X-ray luminosity, while clusters with
non-detection of radio halos should have a radio power much below the
correlation line. These two populations of clusters are found to
correspond to different dynamical states, i.e., clusters with radio
halos showing merging features and those without radio halos being
more relaxed in general \citep[e.g.,][]{CEGB+10}.

Because the X-ray luminosity and gas temperature of galaxy clusters
are tightly related to cluster mass, the relations of $P_{\rm
  1.4~GHz}$\textendash$L_{\rm X}$ and $P_{\rm
  1.4~GHz}$\textendash$T_{\rm X}$ may indicate that emission of halos
and mini-halos is fundamentally related to cluster mass. The
Sunyaev\textendash Zel'dovich (SZ) parameter, indicated as $Y_{\rm
  SZ}$, is a better mass proxy than the X-ray luminosity, since it is
less affected by the cluster dynamics
\citep[e.g.,][]{MHB+05,WSR+08,APP+10,PAA+14}. \citet{Ba+12} found a
tight correlation between the radio power $P_{\rm 1.4~GHz}$ from the
literature and $Y_{\rm SZ}$ from the early Planck SZ catalog. By using
updated SZ data from the Planck mission \citep{PAA+14b} and radio
measurements from the GMRT cluster survey, \citet{CEB+13} confirmed
the scaling relation between the radio power and the cluster SZ
parameter and also the radio bimodality in the radio\textendash SZ
diagram for massive clusters.

There are indications that radio halos, mini-halos, and relics in
galaxy clusters are related to not only cluster masses but also
dynamical states of clusters \citep[e.g.,][]{CCB+15}.  Recently,
\citet{WH+13} found that the offset of radio power from the $P_{\rm
  1.4~GHz}$\textendash$L_{\rm X}$ relation is closely related to the
dynamical parameter $\Gamma$ defined from the optical galaxy
luminosity distributions (see Sect. 2.3). To extend previous studies,
in this paper we search for an empirical fundamental plane among three
sets of quantities: the synchrotron radio power of halos, relics, and
mini-halos, the cluster mass represented by X-ray luminosity or
estimated from gas mass and the SZ effect, and the cluster dynamical
state obtained quantitatively from X-ray or optical data. In Sect.2,
we calculate the observed radio power of halos, relics, and mini-halos
at three frequencies, 1.4~GHz, 610~MHz, and 325~MHz, and collect the
mass proxies of galaxy clusters, $L_{\rm X}$ and $L_{500}$, and also
the SZ-estimated mass $M_{\rm 500,~SZ}$ and the mass $M_{500}$
estimated from gas mass, and obtain the dynamical parameters,
$\Gamma$, $c$, $\omega$, and $P_{3}/P_{0}$, for a large sample of
galaxy clusters with detected radio halos, relics, and mini-halos. In
Section 3, we compare the data scatter around different scaling
relations and then search for the fundamental plane in the
three-dimensional space of these parameters. Conclusions and
discussions are presented in Section 4.

\begin{table*}
\setlength{\tabcolsep}{1mm}
{\footnotesize
\newdimen\digitwidth    
\setbox0=\hbox{\rm0}
\digitwidth=\wd0
\catcode`!=\active
\def!{\kern\digitwidth}
\begin{center}
\caption{Radio Flux and Power for Radio Halos, Relics, and Mini-halos
  from 75 galaxy clusters}
\begin{tabular}{llllcccrrrr}
\hline
\mc{1}{l}{Name}   &\mc{1}{c}{$z$}   &\mc{1}{c}{Type}   &\mc{1}{c}{Size} &\mc{1}{c}{$S_{\rm 1.4~GHz}$} &\mc{1}{c}{$S_{\rm 610~MHz}$} &\mc{1}{c}{$S_{\rm 325~MHz}$} &\mc{1}{c}{References} &\mc{1}{c}{$\log P_{\rm 1.4~GHz}$}  &\mc{1}{c}{$\log P_{\rm 610~MHz}$}  &\mc{1}{c}{$\log P_{\rm 325~MHz}$}  \\
\mc{1}{l}{(1)} &\mc{1}{c}{(2)} &\mc{1}{c}{(3)} &\mc{1}{c}{(4)} &\mc{1}{c}{(5)} &\mc{1}{c}{(6)} &\mc{1}{c}{(7)} &\mc{1}{c}{(8)} &\mc{1}{c}{(9)} &\mc{1}{c}{(10)} &\mc{1}{c}{(11)}\\
\hline
A209                &0.2060   &halo               &$7'$    &15.0$\pm$0.7  &24.0$\pm$3.6  &...             & !1/!2/!--    & 0.24$\pm$0.02   & 0.45$\pm$0.07  &...                \\ 
A399                &0.0718   &halo               &$7'$    &16$\pm$2      &...           &...             & !3/!--/!--   &-0.70$\pm$0.06   &...             &...                \\
A520                &0.1990   &halo               &$5.5'$  &16.7$\pm$0.6  &42$\pm$15     &85$\pm$5        & !4/!0/!4     & 0.26$\pm$0.02   & 0.66$\pm$0.19  & 0.96$\pm$0.03     \\
A521                &0.2533   &halo(+relic)       &$5'$    &6.4$\pm$0.6   &15$\pm$4      &90$\pm$7        & !5/!6/!6     & 0.07$\pm$0.04   & 0.44$\pm$0.13  & 1.22$\pm$0.04     \\
A545                &0.1540   &halo               &$5.6'$  &23$\pm$1      &...           &...             & !7/!--/!--   & 0.15$\pm$0.02   &...             &...                \\[1mm] 
A665                &0.1819   &halo               &$10'$   &43.1$\pm$2.2  &...           &...             & !8/!--/!--   & 0.58$\pm$0.02   &...             &...                \\
A697                &0.2820   &halo               &$2.5'$  &5.2$\pm$0.5   &13.0$\pm$2.0  &47.3$\pm$2.7    & !9/!2/10     & 0.08$\pm$0.04   & 0.48$\pm$0.07  & 1.04$\pm$0.03     \\ 
A746                &0.2320   &halo(+relic)       &$4'$    &18$\pm$4      &...           &...             & !9/!--/!--   & 0.43$\pm$0.11   &...             &...                \\
A754                &0.0542   &halo(+relic)       &$16'$   &83$\pm$5$^a$  &284$\pm$17    &722$\pm$41      & 11/!0/11     &-0.24$\pm$0.03   & 0.29$\pm$0.03  & 0.70$\pm$0.03     \\
A773                &0.2170   &halo               &$6'$    &12.7$\pm$1.3  &...           &...             & 12/!--/!--   & 0.22$\pm$0.05   &...             &...                \\[1mm]
A1300               &0.3072   &halo(+relic)       &$4.8'$  &...           &...           &130$\pm$10      &!--/!--/!1    &...              &...             & 1.56$\pm$0.03     \\
A1351               &0.3224   &halo               &$3'$    &32.4$\pm$4.0  &...           &...             & 13/!--/!--   & 1.01$\pm$0.06   &...             &...                \\ 
A1689               &0.1832   &halo               &$4'$    &9.6$\pm$2.8$^b$&...          &...             & 14/!--/!--   &-0.06$\pm$0.15   &...             &...                \\
A1758N              &0.2790   &halo               &$6'$    &23$\pm$5      &...           & 155$\pm$12     & !1/!--/!1    & 0.72$\pm$0.11   &...             &1.55$\pm$0.03      \\
A1914               &0.1712   &halo               &$7.5'$  &64$\pm$3      &...           &...             & !7/!--/!--   & 0.70$\pm$0.02   &...             &...                \\[1mm]
A1995               &0.3186   &halo               &$3'$    &4.1$\pm$0.7   &...           &...             & 15/!--/!--   & 0.10$\pm$0.08   &...             &...                \\ 
A2069               &0.1160   &halo               &$5'$    &...           &...           &25$\pm$9        &!-- /!--/16   &...              &...             &-0.07$\pm$0.19     \\ 
A2163               &0.2030   &halo               &$11'$   &155$\pm$2     &411$\pm$5     &861$\pm$10      & 17/!0/18     & 1.24$\pm$0.01   & 1.67$\pm$0.01  & 1.99$\pm$0.01     \\
A2218               &0.1756   &halo               &$2'$    &4.7$\pm$0.1   &...           &...             & !8/!--/!--   &-0.41$\pm$0.01   &...             &...                \\
A2219               &0.2256   &halo               &$8'$    &81$\pm$4      &...           &...             & !7/!--/!--   & 1.06$\pm$0.02   &...             &...                \\[1mm] 
A2255               &0.0806   &halo(+relic)       &$10'$   &56$\pm$3      &194$\pm$10    &496$\pm$7       & 19/!0/20     &-0.06$\pm$0.02   & 0.48$\pm$0.02  & 0.89$\pm$0.01     \\
A2256               &0.0581   &halo(+relic)       &$12'$   &103.4$\pm$1.1 &322$\pm$3     &760$\pm$70      & 21/!0/22     &-0.08$\pm$0.01   & 0.41$\pm$0.01  & 0.78$\pm$0.04     \\ 
A2319               &0.0557   &halo               &$16'$   &240$\pm$10    &...           &...             & 23/!--/!--   & 0.24$\pm$0.02   &...             &...                \\ 
A2744               &0.3080   &halo(+relic)       &$7'$    &57$\pm$3      &153$\pm$8     &323$\pm$26      & !1/!0/!1     & 1.21$\pm$0.02   & 1.64$\pm$0.02  & 1.96$\pm$0.04     \\
A3562               &0.0490   &halo               &$5'$    &20$\pm$2      &90$\pm$9      &195$\pm$39      & 24/25/25     &-0.95$\pm$0.05   &-0.30$\pm$0.05  & 0.04$\pm$0.10     \\[1mm]
Bullet              &0.2960   &halo               &$8'$    &56.4$\pm$2.3  &...           &...             & 26/!--/!--   & 1.16$\pm$0.02   &...             &...                \\ 
CL0016+16           &0.5456   &halo               &$2.5'$  &5.5$\pm$0.8$^c$&...          &...             & !8/!--/!--   & 0.76$\pm$0.07   &...             &...                \\
CL0217+70           &0.0655   &halo               &$10'$   &58.6$\pm$0.9  &156$\pm$2     &326$\pm$30      & 27/!0/27     &-0.22$\pm$0.05   & 0.20$\pm$0.01  & 0.52$\pm$0.04     \\
CL1821+643          &0.299    &halo               &$4'$    &14.3$\pm$0.7  &33$\pm$2      &62$\pm$4        & !0/!0/28     & 0.58$\pm$0.02   & 0.94$\pm$0.03  & 1.22$\pm$0.03     \\
Coma                &0.0231   &halo(+relic)       &$30'$   &530$\pm$50    &1200$\pm$300  &3180$\pm$30     & 29/30/31     &-0.19$\pm$0.04   & 0.16$\pm$0.12  & 0.58$\pm$0.01     \\[1mm]
MACS J0553--3342    &0.431    &halo               &$4'$    &...           &...           &62$\pm$5        &!--/!--/32    &...              &...             & 1.57$\pm$0.04     \\
MACS J0717+3745     &0.5458   &halo               &$4'$    &118$\pm$5     &162$\pm$0.23  &337.5$\pm$0.5   &33/34/!--     & 2.09$\pm$0.02   & 2.23$\pm$0.01  & 2.55$\pm$0.01     \\
MACS J1752+4440     &0.366    &halo(+relic)       &$3.3'$  &...           &...           &164$\pm$13      &!--/!--/32    &...              &...             & 1.84$\pm$0.04     \\ 
PLCK G171.9--40.7   &0.270    &halo               &$5.5'$  &18$\pm$2      &...           &...             & 35/!--/!--   & 0.58$\pm$0.05   &...             &...                \\ 
PLCK G287.0+32.9    &0.39     &halo(+relic)       &$4'$    &3.6$\pm$0.5$^c$&26$\pm$4$^c$ &63$\pm$10$^c$    &!0/36/36      & 0.24$\pm$0.06   & 1.10$\pm$0.07  & 1.48$\pm$0.08     \\[1mm] 
RXC J0107+5408      &0.1066   &halo               &$9.5'$  & 55$\pm$5     &...           &...             &!9/!--/!--    & 0.19$\pm$0.04   &...             &...                \\
RXC J1314--2515     &0.2439   &halo(+relic)       &$7'$    &...           &10.3$\pm$0.3  &40$\pm$3        &!--/!2/!1     &...              & 0.24$\pm$0.01  & 0.83$\pm$0.03     \\ 
RXC J1514--1523     &0.2226   &halo               &$7'$    & 10$\pm$2     & 37$\pm$8     &102$\pm$9       &37/!0/37      & 0.14$\pm$0.10   & 0.71$\pm$0.11  & 1.15$\pm$0.04     \\
RXC J2003.5--2323   &0.3173   &halo               &$5'$    & 35$\pm$2     &96.9$\pm$5.0  &235$\pm$12      &38/!2/!0      & 1.03$\pm$0.03   & 1.47$\pm$0.02  & 1.85$\pm$0.02     \\
Toothbrush          &0.225    &halo               &$9'$    &35.9$\pm$2.6  &51.5$\pm$7.4  &121.6$\pm$13.5  &39/39/39      & 0.70$\pm$0.03   & 0.86$\pm$0.07  & 1.23$\pm$0.05     \\[1mm]
Z5247               &0.229    &halo(+relic)       &$4'$    &2$\pm$0.3     &7.9$\pm$1     &...             &40/40/!--     &-0.53$\pm$0.07   & 0.06$\pm$0.06  &...                \\[1mm]
A115                &0.1971   &relic              &$12.5'$ &14.7$\pm$2.2$^c$&...         &...             & 12/!--/!--   & 0.19$\pm$0.07   &...             &...                \\
A521                &0.2533   &relic(+halo)       &$4.2'$  &15.0$\pm$0.8  &41.9$\pm$2.1  &114$\pm$6       & 41/!2/!1     & 0.44$\pm$0.02   & 0.89$\pm$0.02  & 1.32$\pm$0.02     \\
A746                &0.2320   &relic(+halo)       &$5'$    &24.5$\pm$2.0  &...           &...             & !9/!--/!--   & 0.57$\pm$0.04   &...             &...                \\
A754                &0.0542   &relic(+halo)       &$13'$   &6.0$\pm$0.3   &31$\pm$2      &106$\pm$5       & 11/!0/11     &-1.38$\pm$0.02   &-0.67$\pm$0.03  &-0.14$\pm$0.02     \\ 
A1240-N             &0.1590   &relic              &$4'$    & 6.0$\pm$0.2  &12.2$\pm$0.4  &21.0$\pm$0.8    & 42/!0/42     &-0.40$\pm$0.01   &-0.09$\pm$0.01  & 0.14$\pm$0.02     \\
A1240-S             &0.1590   &relic              &$7.5'$  &10.1$\pm$0.4  &18.2$\pm$0.7  &28.5$\pm$1.1    & 42/!0/42     &-0.17$\pm$0.02   & 0.08$\pm$0.02  & 0.28$\pm$0.02     \\[1mm]
A1300               &0.3072   &relic(+halo)       &$2.5'$  &...           &...           &75$\pm$6        &!--/!--/!1    &...              &...             & 1.32$\pm$0.04     \\
A1612               &0.179    &relic              &$4.3'$  &62.8$\pm$2.6  &...           &...             &!9/!--/!--    & 0.73$\pm$0.02   &...             &...                \\
A2061               &0.0784   &relic              &$7.5'$  &27.6$\pm$1.0  &...           &...             &!9/!--/!--    &-0.39$\pm$0.02   &...             &...                \\
A2255               &0.0806   &relic(+halo)       &$8'$    &23$\pm$1      &58$\pm$3      &117$\pm$2       &19/!0/20      &-0.44$\pm$0.02   &-0.04$\pm$0.02  & 0.26$\pm$0.01     \\
A2256-G             &0.0581   &relic(+halo)       &$5'$    &231.6$\pm$15.1&447$\pm$30    &735.7$\pm$45.8  &43/!0/43      & 0.27$\pm$0.03   & 0.55$\pm$0.03  & 0.77$\pm$0.03     \\ 
A2256-H             &0.0581   &relic(+halo)       &$5'$    &245.8$\pm$19.1&475$\pm$37    &781.3$\pm$64.3  &43/!0/43      & 0.29$\pm$0.04   & 0.58$\pm$0.04  & 0.79$\pm$0.04     \\[1mm]
A2345-E             &0.1765   &relic              &$8.5'$  &29.0$\pm$0.4  &84$\pm$1      &188$\pm$3       &42/!0/42      & 0.38$\pm$0.01   & 0.84$\pm$0.01  & 1.19$\pm$0.01     \\
A2345-W             &0.1765   &relic              &$6.5'$  &30.0$\pm$0.5  &109$\pm$2     &291$\pm$4       &42/!0/42      & 0.40$\pm$0.01   & 0.96$\pm$0.01  & 1.38$\pm$0.01     \\
A2744               &0.3080   &relic(+halo)       &$6'$    &20$\pm$1      &54$\pm$8      &122$\pm$10      &!1/!0/!1      & 0.75$\pm$0.02   & 1.18$\pm$0.07  & 1.54$\pm$0.04     \\ 
A3365-E             &0.0926   &relic              &$5.5'$  &42.6$\pm$2.6  &...           &...             &!9/!--/!--    &-0.05$\pm$0.03   &...             &...                \\
A3365-W             &0.0926   &relic              &$2.3'$  &5.3$\pm$0.5   &...           &...             &!9/!--/!--    &-0.95$\pm$0.01   &...             &...                \\
A3376-E             &0.0456   &relic              &$16'$   &122$\pm$10    &559$\pm$46    &1770$\pm$90     &44/!0/44      &-0.23$\pm$0.04   & 0.43$\pm$0.04  & 0.93$\pm$0.02     \\
A3376-W             &0.0456   &relic              &$15'$   &113$\pm$10    &467$\pm$41    &1367$\pm$70     &44/!0/44      &-0.26$\pm$0.04   & 0.35$\pm$0.04  & 0.82$\pm$0.02     \\ 
A3667-SE            &0.0556   &relic              &$20'$   &350$\pm$20     &...          &...             &45/!--/!--    & 0.41$\pm$0.03   &...             &...                \\ 
A3667-NW            &0.0556   &relic              &$30'$   &2470$\pm$170   &...          &...             &45/!--/!--    & 1.25$\pm$0.03   &...             &...                \\
\hline
\end{tabular}
\label{tab1}
\end{center}}
\end{table*}

\addtocounter{table}{-1}
\begin{table*}[t]
\setlength{\tabcolsep}{0.5mm}
{\footnotesize
\newdimen\digitwidth    
\setbox0=\hbox{\rm0}
\digitwidth=\wd0
\catcode`!=\active
\def!{\kern\digitwidth}
\begin{center}
\caption{{\it continued}}
\begin{tabular}{llllcccrrrr}
\hline
\mc{1}{l}{Name}   &\mc{1}{c}{$z$}   &\mc{1}{c}{Type}   &\mc{1}{c}{Size} &\mc{1}{c}{$S_{\rm 1.4~GHz}$} &\mc{1}{c}{$S_{\rm 610~MHz}$} &\mc{1}{c}{$S_{\rm 325~MHz}$} &\mc{1}{c}{References} &\mc{1}{c}{$\log P_{\rm 1.4~GHz}$}  &\mc{1}{c}{$\log P_{\rm 610~MHz}$}  &\mc{1}{c}{$\log P_{\rm 325~MHz}$}  \\
\mc{1}{l}{(1)} &\mc{1}{c}{(2)} &\mc{1}{c}{(3)} &\mc{1}{c}{(4)} &\mc{1}{c}{(5)} &\mc{1}{c}{(6)} &\mc{1}{c}{(7)} &\mc{1}{c}{(8)} &\mc{1}{c}{(9)} &\mc{1}{c}{(10)} &\mc{1}{c}{(11)} \\
\hline
CIZA J0649+1801     &0.064     &relic             &$11'$   &...            &321$\pm$46   &...              &!--/!9/!--    &...             & 0.49$\pm$0.07  &...                \\
CIZA J2242+5301-N   &0.1921    &relic(+halo)      &$9'$    &144$\pm$15     &...          &...              &46/!--/!--    & 1.16$\pm$0.05  &...             &...                \\ 
CIZA J2242+5301-S   &0.1921    &relic(+halo)      &$7.5'$  &18$\pm$2       &...          &...              &46/!--/!--    & 0.25$\pm$0.05  &...             &...                \\
Coma                &0.0231    &relic(+halo)      &$30'$   &260$\pm$39$^c$ &...          &...              &47/!--/!--    &-0.50$\pm$0.07  &...             &...                \\
El Gordo-NW         &0.870     &relic             &$1'$    &7.0$\pm$0.5    &19$\pm$2     &...              &!0/48/!--     & 1.33$\pm$0.03  & 1.77$\pm$0.05  &...                \\ 
MACS J1752+4440-NE  &0.366     &relic(+halo)      &$4'$    &65.3$\pm$3.9   &...          &410$\pm$33       &49/!--/32     & 1.44$\pm$0.03  &...             & 2.23$\pm$0.04     \\
MACS J1752+4440-SW  &0.366     &relic(+halo)      &$3'$    &30.2$\pm$1.8   &...          &163$\pm$13       &49/!--/32     & 1.10$\pm$0.03  &...             & 1.83$\pm$0.04     \\[1mm]
PLCK G287.0+32.9-NW &0.39      &relic(+halo)      &$4.5'$  &27$\pm$4$^c$   &110$\pm$11$^c$&216$\pm$32$^c$   &!0/36/36      & 1.12$\pm$0.07  & 1.73$\pm$0.05  & 2.02$\pm$0.07     \\
PLCK G287.0+32.9-SE &0.39      &relic(+halo)      &$3.5'$  &10$\pm$2$^c$   &50$\pm$5$^c$  &114$\pm$17$^c$   &!0/36/36      & 0.68$\pm$0.10  & 1.38$\pm$0.05  & 1.60$\pm$0.07     \\
PSZ1 G096.9+24.2-N  &0.3       &relic             &$3.3'$  &8.9$\pm$0.8    &...          &...              &50/!--/!--    & 0.38$\pm$0.21  &...             &...                \\
PSZ1 G096.9+24.2-S  &0.3       &relic             &$5.3'$  &18.3$\pm$1.9   &...          &...              &50/!--/!--    & 0.69$\pm$0.05  &...             &...                \\
RXC J1053+5452      &0.0704    &relic             &$7.5'$  &15$\pm$2       &...          &...              &!9/!--/!--    &-0.75$\pm$0.06  &...             &...                \\
RXC J1314--2515-E   &0.2439    &relic(+halo)      &$2.5'$  &10.1$\pm$0.3   &28.0$\pm$1.4 &52$\pm$4         &51/!2/!1      & 0.23$\pm$0.01  & 0.69$\pm$0.02  & 0.94$\pm$0.03     \\
RXC J1314--2515-W   &0.2439    &relic(+halo)      &$4.5'$  &20.2$\pm$0.5   &64.8$\pm$3.2 &137$\pm$11       &51/!2/!1      & 0.53$\pm$0.01  & 1.04$\pm$0.02  & 1.36$\pm$0.04     \\
Toothbrush          &0.225     &relic             &$8.5'$  &319.5$\pm$20.8 &797$\pm$52   &1600$\pm$100     &39/39/59      & 1.65$\pm$0.03  & 2.05$\pm$0.03  & 2.35$\pm$0.03     \\[1mm]
Z5247               &0.229     &relic(+halo)      &$3'$    &3.1$\pm$0.2    &9.3$\pm$1.0  &23.1$\pm$2.5     &40/40/!0      &-0.34$\pm$0.03  & 0.13$\pm$0.05  & 0.53$\pm$0.05     \\
ZwCl0008+5215-E     &0.1032    &relic             &$12'$   &56.0$\pm$3.5   &230$\pm$25   &545$\pm$59       &52/52/!0      & 0.17$\pm$0.03  & 0.78$\pm$0.05  & 1.16$\pm$0.05     \\ 
ZwCl0008+5215-W     &0.1032    &relic             &$2.5'$  &11.0$\pm$1.2   &56$\pm$8     &89$\pm$13        &52/52/!0      &-0.54$\pm$0.05  & 0.17$\pm$0.07  & 0.37$\pm$0.07     \\
ZwCl2341+0000-N     &0.27      &relic             &$1'$    &...            &14$\pm$3     &18$\pm$4         &!--/53/!0     &...             & 0.47$\pm$0.10  & 0.58$\pm$0.11     \\ 
ZwCl2341+0000-S     &0.27      &relic             &$5'$    &...            &37$\pm$13    &58$\pm$20        &!--/53/!0     &...             & 0.89$\pm$0.19  & 1.09$\pm$0.18     \\[1mm]
A478                &0.088     &mini-halo         &$3'$    &16.6$\pm$3     &...          &...              &54/!--/!--    &-0.50$\pm$0.09  &...             &...                \\
A1835               &0.2532    &mini-halo         &$2'$    &6.1$\pm$1.3    &...          &...              &54/!--/!--    & 0.05$\pm$0.10  &...             &...                \\ 
A2029               &0.0765    &mini-halo         &$6'$    &19.5$\pm$2.5   &...          &...              &54/!--/!--    &-0.56$\pm$0.06  &...             &...                \\
A2204               &0.152     &mini-halo         &$0.6'$  &8.6$\pm$0.9    &...          &...              &54/!--/!--    &-0.29$\pm$0.05  &...             &...                \\
A2390               &0.228     &mini-halo         &$2'$    &28.3$\pm$4.3   &...          &...              &54/!--/!--    & 0.61$\pm$0.07  &...             &...                \\[1mm]
A3444               &0.254     &mini-halo         &$0.5'$  &...            &29.5$\pm$0.5 &...              &!--/40/!--    &...             & 0.74$\pm$0.01  &...                \\
2A0335              &0.0347    &mini-halo         &$3.5'$  &21.1$\pm$2.1   &...          &...              &54/!--/!--    &-1.23$\pm$0.05  &...             &...                \\
MS 1455+2232        &0.2578    &mini-halo         &$2'$    &8.5$\pm$1.1    &...          &...              &54/!--/!--    & 0.21$\pm$0.06  &...             &...                \\ 
Ophiuchus           &0.028     &mini-halo         &$15'$   &83.4$\pm$6.6   &...          &...              &54/!--/!--    &-0.83$\pm$0.04  &...             &...                \\
Perseus             &0.0179    &mini-halo         &$12'$   &3020$\pm$153   &...          &...              &54/!--/!--    & 0.34$\pm$0.02  &...             &...                \\[1mm]
Phoenix             &0.596     &mini-halo         &$1'$    &...            &17$\pm$5     &...              &!--/55/!--    &...             & 1.34$\pm$0.15  &...                \\
RBS797              &0.35      &mini-halo         &$1'$    &5.2$\pm$0.6    &...          &...              &54/!--/!--    & 0.29$\pm$0.05  &...             &...                \\ 
RXC J1504--0248     &0.2153    &mini-halo         &$1'$    &20.0$\pm$1.0   &59$\pm$3     &121$\pm$6        &54/40/56      & 0.41$\pm$0.02  & 0.88$\pm$0.02  & 1.19$\pm$0.02     \\
RXC J1532+3021      &0.3621    &mini-halo         &$0.7'$  &7.5$\pm$0.4    &16$\pm$1     &33.5$\pm$4.4     &54/54/54      & 0.49$\pm$0.02  & 0.81$\pm$0.03  & 1.14$\pm$0.06     \\ 
RX J1347--1145      &0.4516    &mini-halo         &$2'$    &34.1$\pm$2.3   &...          &...              &54/!--/!--    & 1.36$\pm$0.03  &...             &...                \\[1mm]
RX J1720+2638       &0.1644    &mini-halo         &$1.5'$  &68$\pm$5       &170$\pm$12   &365$\pm$58       &57/57/57      & 0.69$\pm$0.03  & 1.08$\pm$0.03  & 1.41$\pm$0.08     \\
RX J2129+0005       &0.235     &mini-halo         &$1'$    &2.4$\pm$0.4$^c$&8$\pm$1      &...              &40/40/!--     &-0.43$\pm$0.08  & 0.09$\pm$0.06  &...                \\
S780                &0.236     &mini-halo         &$0.5'$  &...            &35$\pm$9     &...              &!--/40/!--    &...             & 0.74$\pm$0.13  &...                \\
Z3146               &0.290     &mini-halo         &$1'$    &5.2$\pm$0.8$^c$&...          &...              &54/!--/!--    & 0.11$\pm$0.07  &...             &...                \\ 
\hline
\end{tabular}
\label{tab1}
\end{center}
}
{Notes: Columns: (1) cluster name; (2) redshift; (3) type of diffuse radio emission\textemdash halos, relics, or mini-halos. Known ``radio Phoenix'' looks like it is from a radio galaxy and is not included in this table; (4) angular size; (5)\textendash(7) flux of halos, relics, and mini-halos at 1.4 GHz, 610 MHz, and 325 MHz, all in mJy; (8) reference numbers of these radio fluxes; (9)\textendash(11) radio power (in $10^{24}$~W/Hz) at these three frequencies after the {\it k}-correction.\\
Notes for some measurements are:\\
$^a$ Caculated from fluxes of whole emission region minus relic region.\\
$^b$ Estimated from flux at a nearby frequency.\\
$^c$ Uncertainty assumed to be 15\%.\\
References. 0\textemdash estimated by us based on measurements available at other frequencies; 1\textemdash \citet{VGDC+13}; 2\textemdash \citet{VGDC+08}; 3\textemdash \citet{MGFG+10}; 4\textemdash \citet{VFG+14}; 5\textemdash \citet{DBG+09}; 6\textemdash \citet{BGCL+08}; 7\textemdash \citet{BFG+03}; 8\textemdash \citet{GF+00}; 9\textemdash \citet{VBR+11}; 10\textemdash \citet{MVBD+10}; 11\textemdash \citet{MMG+11}; 12\textemdash \citet{GFG+01}; 13\textemdash \citet{GVC+09}; 14\textemdash \citet{VGM+11}; 15\textemdash \citet{GBF+09}; 16\textemdash \citet{DHP+15}; 17\textemdash \citet{FFGG+01}; 18\textemdash \citet{FOBG+04}; 19\textemdash \citet{GMF+05}; 20\textemdash \citet{PD09}; 21\textemdash \citet{CE+06}; 22\textemdash \citet{Br08}; 23\textemdash \citet{SJR15}; 24\textemdash \citet{VBDB+03}; 25\textemdash \citet{GVBB+05}; 26\textemdash \citet{SBF+14}; 27\textemdash \citet{BDR+11}; 28\textemdash \citet{BIB+14}; 29\textemdash \citet{KKD+90}; 30\textemdash \citet{GFV+93}; 31\textemdash \citet{VGF+90}; 32\textemdash \citet{BBVV+12}; 33\textemdash \citet{BFGG+09}; 34\textemdash \citet{PRC+13}; 35\textemdash \citet{GKWV+13}; 36\textemdash \citet{BIBG+14}; 37\textemdash \citet{GDVB+11}; 38\textemdash \citet{GVBD+09}; 39\textemdash \citet{VRI+12}; 40\textemdash \citet{KVG+15}; 41\textemdash \citet{GVM+08}; 42\textemdash \citet{BGFG+09}; 43\textemdash \citet{TAL+15}; 44\textemdash \citet{KDB+12}; 45\textemdash \citet{RSO+15}; 46\textemdash \citet{VBRH+11}; 47\textemdash \citet{GFS+91}; 48\textemdash \citet{LBH+14}; 49\textemdash \citet{VBE+12}; 50\textemdash \citet{DVB+14}; 51\textemdash \citet{FSBG+05}; 52\textemdash \citet{VHR+11}; 53\textemdash \citet{VRB+09}; 54\textemdash \citet{GMV+14}; 55\textemdash \citet{VILA+14}; 56\textemdash \citet{GMB+11}; 57\textemdash \citet{GMB+14}.}
\vspace{-20mm}
\end{table*}

\begin{table*}
\setlength{\tabcolsep}{0.5mm}
{\footnotesize
\newdimen\digitwidth    
\setbox0=\hbox{\rm0}
\digitwidth=\wd0
\catcode`!=\active
\def!{\kern\digitwidth}
\begin{center}
\caption{Masses, Mass Proxies, and Dynamical Parameters for 75 galaxy clusters}
\begin{tabular}{lcccccrrrrr}
\hline
\mc{1}{l}{Name} &\mc{1}{c}{$\log L_{\rm X}$} &\mc{1}{c}{$\log L_{500}$} &\mc{1}{c}{$\log M_{\rm SZ,~500}$}  &\mc{1}{c}{$\log M_{500}$} &\mc{1}{c}{References} &\mc{1}{c}{$\Gamma$} &\mc{1}{c}{$\log c$} &\mc{1}{c}{$\log \omega$} &\mc{1}{c}{$\log (P_{3}/P_{0})$}  &\mc{1}{c}{References}\\
\mc{1}{l}{(1)} &\mc{1}{c}{(2)} &\mc{1}{c}{(3)} &\mc{1}{c}{(4)} &\mc{1}{c}{(5)} &\mc{1}{c}{(6)} &\mc{1}{c}{(7)} &\mc{1}{c}{(8)} &\mc{1}{c}{(9)} &\mc{1}{c}{(10)} &\mc{1}{c}{(11)}\\
\hline
A115             & 0.96$\pm$0.12  & 0.87$\pm$0.12   & 0.88$\pm$0.02   &...             &!1/!2/!3/!-- &$-$0.76$\pm$0.11 &$-$0.49$\pm$0.01 &$-$2.34$\pm$0.07 &$-$6.06$\pm$0.06 &28/!0/!0/!0      \\ 
A209             & 0.80$\pm$0.05  & 0.88$\pm$0.01   & 0.93$\pm$0.02   & 1.03$\pm$0.07  &!4/!5/!3/!6  &$-$0.24$\pm$0.11 &$-$0.72$\pm$0.02 &$-$1.60$\pm$0.02 &$-$7.27$\pm$0.54 &!0/29/!0/!0      \\ 
A399             & 0.59$\pm$0.06  & 0.26$\pm$0.01   & 0.72$\pm$0.02   & 0.76$\pm$0.02  &!1/!7/!3/!6  &   0.13$\pm$0.05 &...              &...              &...              &!0/!--/!--/!--   \\ 
A478             & 0.87$\pm$0.02  & 1.01$\pm$0.01   & 0.84$\pm$0.02   & 0.91$\pm$0.02  &!8/!7/!3/!6  &...              &...              &...              &...              &!--/!--/!--/!--  \\
A520             & 0.95$\pm$0.11  & 0.89$\pm$0.01   & 0.89$\pm$0.02   & 1.01$\pm$0.06  &!1/!5/!3/!6  &$-$0.27$\pm$0.06 &$-$1.04$\pm$0.01 &$-$0.80$\pm$0.01 &$-$6.05$\pm$0.03 &!0/29/!0/!0      \\[1.1mm]
A521             & 0.91$\pm$0.08  & 0.92$\pm$0.01   & 0.86$\pm$0.03   & 0.99$\pm$0.07  &!4/!5/!3/!6  &...              &$-$1.01$\pm$0.02 &$-$1.12$\pm$0.01 &$-$5.87$\pm$0.26 &!--/29/!0/!0     \\ 
A545             & 0.75$\pm$0.04  & 0.80$\pm$0.01   & 0.73$\pm$0.03   &...             &!4/!5/!3/!-- &$-$0.35$\pm$0.04 &...              &...              &...              &!0/!--/!--/!--   \\ 
A665             & 0.99$\pm$0.07  & 0.92$\pm$0.01   & 0.95$\pm$0.02   & 1.04$\pm$0.07  &!1/!5/!3/!6  &$-$0.26$\pm$0.10 &$-$0.76$\pm$0.01 &$-$1.11$\pm$0.01 &$-$6.48$\pm$0.03 &28/!0/!0/!0      \\
A697             & 1.02$\pm$0.08  & 1.11$\pm$0.01   & 1.04$\pm$0.01   & 1.16$\pm$0.08  &!9/!5/!3/!6  &$-$0.22$\pm$0.06 &$-$0.82$\pm$0.02 &$-$1.93$\pm$0.09 &$-$6.77$\pm$0.30 &28/29/!0/!0      \\ 
A746             & 0.57$\pm$0.15* & 0.53$\pm$0.15   & 0.73$\pm$0.03   &...             &10/!0/!3/!-- &$-$2.49$\pm$0.10 &...              &...              &...              &!0/!--/!--/!--   \\[1.1mm] 
A754             & 0.63$\pm$0.01  & 0.27$\pm$0.01   & 0.84$\pm$0.01   & 0.84$\pm$0.02  &!4/!7/!3/!6  &$-$0.17$\pm$0.11 &...              &...              &...              &!0/!--/!--/!--   \\
A773             & 0.91$\pm$0.08  & 0.86$\pm$0.01   & 0.84$\pm$0.02   & 0.87$\pm$0.06  &!1/!5/!3/!6  &$-$0.12$\pm$0.07 &$-$0.74$\pm$0.01 &$-$1.56$\pm$0.01 &$-$6.97$\pm$0.09 &28/29/!0/!0      \\
A1240            &-0.01$\pm$0.02  &-0.05$\pm$0.03   &...              &...             &11/!0/!--/!--&$-$0.52$\pm$0.12 &...              &...              &...              &28/!--/!--/!--   \\
A1300            & 1.15$\pm$0.07  & 1.06$\pm$0.01   & 0.95$\pm$0.02   & 1.27$\pm$0.07  &!4/!5/!3/!6  &$-$0.95$\pm$0.12 &$-$0.72$\pm$0.02 &$-$1.31$\pm$0.01 &$-$6.07$\pm$0.08 &!0/29/!0/!0      \\ 
A1351            & 0.74$\pm$0.11  & 0.72$\pm$0.11   & 0.84$\pm$0.02   &...             &!8/!2/!3/!-- &$-$1.23$\pm$0.15 &$-$1.07$\pm$0.02 &$-$1.18$\pm$0.01 &$-$6.24$\pm$0.26 &28/!0/!0/!0      \\[1.1mm] 
A1612            & 0.39$\pm$0.18  & 0.38$\pm$0.18   & 0.65$\pm$0.05   &...             &!4/!2/!3/!-- &$-$1.71$\pm$0.12 &...              &...              &...              &28/!--/!--/!--   \\ 
A1689            & 1.15$\pm$0.04  & 1.16$\pm$0.01   & 0.94$\pm$0.02   & 0.96$\pm$0.07  &!4/!7/!3/!6  &   0.47$\pm$0.05 &$-$0.46$\pm$0.01 &$-$2.65$\pm$0.77 &$-$7.96$\pm$0.14 &28/!0/!0/!0      \\ 
A1758N           & 1.09$\pm$0.10  & 0.94$\pm$0.01   & 0.91$\pm$0.02   &...             &12/!5/!3/!-- &$-$0.70$\pm$0.08 &$-$0.99$\pm$0.01 &$-$0.84$\pm$0.01 &$-$5.35$\pm$0.03 &28/29/!0/!0      \\ 
A1835            & 1.39$\pm$0.06  & 1.38$\pm$0.01   & 0.99$\pm$0.02   & 1.02$\pm$0.05  &!1/!7/!3/!6  &   0.56$\pm$0.02 &$-$0.43$\pm$0.02 &$-$2.55$\pm$0.37 &$-$8.10$\pm$0.50 &28/!0/!0/!0      \\ 
A1914            & 1.03$\pm$0.04  & 0.98$\pm$0.01   & 0.86$\pm$0.02   & 0.96$\pm$0.07  &12/!7/!3/!6  &$-$0.36$\pm$0.10 &$-$0.65$\pm$0.01 &$-$1.17$\pm$0.01 &$-$6.95$\pm$0.04 &28/!0/!0/!0      \\[1.1mm]
A1995            & 0.95$\pm$0.06  & 0.78$\pm$0.01   & 0.69$\pm$0.03   &...             &!8/!5/!3/!-- &$-$0.09$\pm$0.07 &...              &...              &...              &28/!--/!--/!--   \\
A2029            & 0.95$\pm$0.12  & 0.89$\pm$0.01   & 0.85$\pm$0.01   & 0.93$\pm$0.02  &12/!7/!3/!6  &   0.40$\pm$0.03 &$-$0.37$\pm$0.01 &$-$2.50$\pm$0.01 &$-$9.28$\pm$0.54 &28/!0/!0/!0      \\ 
A2061            & 0.31$\pm$0.07  & 0.27$\pm$0.07   & 0.56$\pm$0.03   &...             &!1/!2/!3/!-- &$-$0.58$\pm$0.11 &...              &...              &...              &28/!--/!--/!--   \\
A2069            & 0.66$\pm$0.07  & 0.63$\pm$0.07   & 0.73$\pm$0.02   &...             &!1/!2/!3/!-- &$-$0.26$\pm$0.04 &...              &...              &...              &28/!--/!--/!--   \\
A2163            & 1.36$\pm$0.03  & 1.40$\pm$0.02   & 1.21$\pm$0.01   & 1.41$\pm$0.02  &!4/!7/!3/!6  &$-$1.05$\pm$0.05 &$-$0.90$\pm$0.02 &$-$1.27$\pm$0.01 &$-$6.02$\pm$0.28 &!0/29/!0/!0      \\[1.1mm]
A2204            & 1.14$\pm$0.02  & 1.20$\pm$0.01   & 0.89$\pm$0.02   & 0.98$\pm$0.02  &!8/!7/!3/!6  &   0.28$\pm$0.05 &$-$0.30$\pm$0.01 &$-$3.24$\pm$0.60 &$-$8.90$\pm$0.20 &28/!0/!0/!0      \\ 
A2218            & 0.75$\pm$0.04  & 0.71$\pm$0.04   & 0.82$\pm$0.01   & 0.79$\pm$0.08  &!1/14/!3/!6  &   0.33$\pm$0.03 &$-$0.73$\pm$0.01 &$-$1.82$\pm$0.01 &$-$6.90$\pm$0.08 &!0/!0/!0/!0      \\
A2219            & 1.10$\pm$0.05  & 1.23$\pm$0.01   & 1.07$\pm$0.01   & 1.21$\pm$0.06  &!1/!7/!3/!6  &$-$0.24$\pm$0.06 &$-$0.86$\pm$0.01 &$-$1.58$\pm$0.01 &$-$6.39$\pm$0.04 &!0/29/!0/!0      \\
A2255            & 0.42$\pm$0.02  & 0.70$\pm$0.05   & 0.73$\pm$0.01   & 0.71$\pm$0.08  &12/!7/!3/!6  &$-$1.02$\pm$0.10 &...              &...              &...              &28/!--/!--/!--   \\
A2256            & 0.58$\pm$0.02  & 0.48$\pm$0.01   & 0.79$\pm$0.01   & 0.80$\pm$0.02  &12/!7/!3/!6  &$-$0.17$\pm$0.08 &...              &...              &...              &28/!--/!--/!--   \\[1.1mm] 
A2319            & 0.87$\pm$0.02  & 0.19$\pm$0.03   & 0.94$\pm$0.01   &...             &13/!7/!3/!-- &   0.14$\pm$0.08 &...              &...              &...              &!0/!--/!--/!--   \\ 
A2345            & 0.63$\pm$0.06  & 0.59$\pm$0.06   & 0.77$\pm$0.03   &...             &!4/!2/!3/!-- &...              &$-$1.17$\pm$0.04 &$-$0.91$\pm$0.01 &$-$5.85$\pm$0.05 &!--/!0/!0/!0     \\
A2390            & 1.13$\pm$0.12  & 1.30$\pm$0.01   & 0.99$\pm$0.01   & 1.11$\pm$0.06  &!1/!7/!3/!6  &   0.04$\pm$0.06 &$-$0.54$\pm$0.01 &$-$2.38$\pm$0.21 &$-$7.24$\pm$0.14 &28/29/!0/!0      \\ 
A2744            & 1.10$\pm$0.05  & 1.17$\pm$0.01   & 0.99$\pm$0.02   & 1.18$\pm$0.06  &!4/!5/!3/!6  &$-$1.03$\pm$0.04 &$-$1.00$\pm$0.02 &$-$1.17$\pm$0.01 &$-$5.91$\pm$0.11 &!0/29/!0/!0      \\ 
A3365            &-0.06$\pm$0.15* &-0.10$\pm$0.15   &...              &...             &15/!0/!--/!--&...              &...              &...              &...              &!--/!--/!--/!--  \\[1.1mm] 
A3376            & 0.03$\pm$0.02  & 0.00$\pm$0.02   & 0.38$\pm$0.03   & 0.41$\pm$0.02  &!4/!2/!3/!6  &...              &...              &...              &...              &!--/!--/!--/!--  \\
A3444            & 1.14$\pm$0.04  & 1.08$\pm$0.04   & 0.87$\pm$0.02   &...             &!4/!2/!3/!-- &...              &$-$0.35$\pm$0.01 &$-$3.27$\pm$0.46 &$-$7.53$\pm$0.03 &!--/!0/!0/!0     \\
A3562            & 0.17$\pm$0.03  & 0.02$\pm$0.01   & 0.39$\pm$0.04   & 0.55$\pm$0.02  &!4/!7/!3/!6  &   0.29$\pm$0.04 &...              &...              &...              &!0/!--/!--/!--   \\
A3667            & 1.05$\pm$0.02  & 0.81$\pm$0.01   & 0.85$\pm$0.01   & 0.95$\pm$0.02  &!4/!7/!3/!6  &...              &...              &...              &...              &!--/!--/!--/!--  \\ 
2A0335           & 0.35$\pm$0.01  & 0.32$\pm$0.01   & 0.36$\pm$0.03   & 0.41$\pm$0.02  &!8/!7/!3/!6  &...              &...              &...              &...              &!--/!--/!--/!--  \\[1.1mm] 
Bullet           & 1.36$\pm$0.04  & 1.35$\pm$0.01   & 1.12$\pm$0.01   & 1.29$\pm$0.06  &!4/!5/!3/!6  &...              &$-$0.90$\pm$0.01 &$-$0.77$\pm$0.01 &$-$5.41$\pm$0.01 &!--/!0/!0/!0     \\
CIZA J0649+1801  & 0.08$\pm$0.06  & 0.07$\pm$0.06   &...              &...             &16/!2/!--/!--&...              &...              &...              &...              &!--/!--/!--/!--  \\
CIZA J2242+5301  & 0.83$\pm$0.10  & 0.58$\pm$0.10   &...              &...             &17/!2/!--/!--&...              &...              &...              &...              &!--/!--/!--/!--  \\ 
CL0016+16        & 1.29$\pm$0.01  & 1.19$\pm$0.01   & 0.99$\pm$0.02   & 1.15$\pm$0.07  &18/!5/!3/!6  &...              &$-$0.85$\pm$0.01 &$-$1.78$\pm$0.06 &$-$6.98$\pm$0.19 &!--/!0/!0/!0     \\
CL0217+70        &-0.20$\pm$0.15* &-0.24$\pm$0.15   &...              &...             &19/!0/!--/!--&...              &...              &...              &...              &!--/!--/!--/!--  \\[1.1mm]
CL1821+64        & 1.16$\pm$0.01  & 1.12$\pm$0.01   & 0.83$\pm$0.02   &...             &20/!0/!3/!-- &...              &$-$0.41$\pm$0.01 &$-$2.64$\pm$0.06 &$-$7.21$\pm$0.08 &!--/!0/!0/!0     \\
Coma             & 0.58$\pm$0.01  & 0.07$\pm$0.01   & 0.86$\pm$0.01   &...             &!1/!7/!3/!-- &$-$0.22$\pm$0.05 &...              &...              &...              &!0/!--/!--/!--   \\
El Gordo         &...             & 1.55$\pm$0.02   & 1.03$\pm$0.02   &...             &!--/21/!3/!--&...              &$-$0.70$\pm$0.01 &$-$0.97$\pm$0.01 &$-$5.46$\pm$0.06 &!--/!0/!0/!0     \\ 
MACS J0553--3342 &...             & 1.23$\pm$0.15   & 0.94$\pm$0.02   &...             &!--/22/!3/!--&...              &$-$0.90$\pm$0.01 &$-$0.91$\pm$0.01 &$-$5.34$\pm$0.02 &!--/!0/!0/!0     \\ 
MACS J0717+3745  & 1.39$\pm$0.01  & 1.38$\pm$0.01   & 1.06$\pm$0.02   & 1.33$\pm$0.05  &18/!5/!3/!6  &...              &$-$0.96$\pm$0.02 &$-$1.79$\pm$0.09 &$-$5.38$\pm$0.04 &!--/!0/!0/!0     \\[1.1mm]
MACS J1752+4440  & 0.92$\pm$0.15* & 0.88$\pm$0.16   & 0.83$\pm$0.03   &...             &23/!0/!3/!-- &$-$1.82$\pm$0.12 &...              &...              &...              &!0/!--/!--/!--   \\
MS1455+2232      & 0.92$\pm$0.12  & 0.94$\pm$0.13   &...              &...             &!8/!2/!--/!--&   0.22$\pm$0.04 &$-$0.24$\pm$0.02 &$-$2.43$\pm$0.18 &$-$7.71$\pm$0.14 &28/29/!0/!0      \\
Ophiuchus        & 0.72$\pm$0.01  & 0.58$\pm$0.01   &...              &...             &16/!7/!--/!--&   0.09$\pm$0.08 &...              &...              &...              &!0/!--/!--/!--   \\
Perseus          & 0.89$\pm$0.01  & 0.79$\pm$0.01   &...              &...             &12/!2/!--/!--&   0.06$\pm$0.06 &...              &...              &...              &!0/!--/!--/!--   \\
Phoenix          &...             &...              & 0.95$\pm$0.03   &...             &!--/!--/!3/!--&...             &$-$0.25$\pm$0.02 &$-$2.63$\pm$0.64 &$-$7.91$\pm$0.66 &!--/!0/!0/!0     \\[1.1mm]
PLCK G171.9--40.7&...             & 1.05$\pm$0.01   & 1.03$\pm$0.02   &...             &!--/!5/!3/!--&...              &$-$0.83$\pm$0.01 &$-$1.75$\pm$0.03 &$-$6.88$\pm$0.09 &!--/!0/!0/!0     \\
PLCK G287.0+32.9 &...             & 1.24$\pm$0.01   & 1.17$\pm$0.01   &...             &!--/24/!3/!--&...              &...              &...              &...              &!--/!--/!--/!--  \\
PSZ1 G096.9+24.2 & 0.58$\pm$0.15* & 0.54$\pm$0.15   & 0.67$\pm$0.03   &...             &25/!0/!3/!-- &...              &...              &...              &...              &!--/!--/!--/!--  \\ 
RBS797           & 1.31$\pm$0.02  & 1.30$\pm$0.02   & 0.75$\pm$0.04   & 0.86$\pm$0.07  &26/!2/!3/!6  &...              &$-$0.25$\pm$0.01 &$-$3.44$\pm$0.23 &$-$9.49$\pm$0.53 &!--/!0/!0/!0     \\ 
RXC J0107+5408   & 0.44$\pm$0.08  & 0.45$\pm$0.08   & 0.77$\pm$0.02   &...             &16/!2/!3/!-- &...              &...              &...              &...              &!--/!--/!--/!--  \\
\hline
\end{tabular}
\end{center}
}
\label{tab2}
\end{table*} 

\addtocounter{table}{-1}
\setlength{\tabcolsep}{0.5mm}
\begin{table*}[htb]
{\footnotesize
\newdimen\digitwidth    
\setbox0=\hbox{\rm0}
\digitwidth=\wd0
\catcode`!=\active
\def!{\kern\digitwidth}
\begin{center}
\caption{{\it continued}}
\begin{tabular}{lcccccrrrrr}
\hline
\mc{1}{l}{Name} &\mc{1}{c}{$\log L_{\rm X}$} &\mc{1}{c}{$\log L_{500}$} &\mc{1}{c}{$\log M_{\rm SZ,~500}$}  &\mc{1}{c}{$\log M_{500}$} &\mc{1}{c}{References} &\mc{1}{c}{$\Gamma$} &\mc{1}{c}{$\log c$} &\mc{1}{c}{$\log \omega$} &\mc{1}{c}{$\log (P_{3}/P_{0}$)}  &\mc{1}{c}{References}\\
\mc{1}{l}{(1)} &\mc{1}{c}{(2)} &\mc{1}{c}{(3)} &\mc{1}{c}{(4)} &\mc{1}{c}{(5)} &\mc{1}{c}{(6)} &\mc{1}{c}{(7)} &\mc{1}{c}{(8)} &\mc{1}{c}{(9)} &\mc{1}{c}{(10)} &\mc{1}{c}{(11)}\\
\hline
RXC J1053+5452   & 0.58$\pm$0.05  &-0.35$\pm$0.05   &...              &...             &!9/!2/!--/!--&...              &$-$0.91$\pm$0.01 &$-$1.26$\pm$0.01 &$-$6.71$\pm$0.16 &!--/!0/!0/!0     \\ 
RXC J1314--2515  & 1.04$\pm$0.08  & 1.00$\pm$0.08   & 0.83$\pm$0.04   &...             &!4/!2/!3/!-- &$-$0.39$\pm$0.07 &...              &...              &...              &!0/!--/!--/!--   \\ 
RXC J1504--0248  & 1.45$\pm$0.02  & 1.45$\pm$0.01   & 0.82$\pm$0.03   &...             &!4/!7/!3/!-- &   0.32$\pm$0.04 &$-$0.22$\pm$0.01 &$-$2.78$\pm$1.08 &$-$8.09$\pm$0.09 &28/29/!0/!0      \\ 
RXC J1514--1523  & 0.85$\pm$0.08  & 0.81$\pm$0.08   & 0.95$\pm$0.02   &...             &!4/!2/!3/!-- &...              &$-$1.19$\pm$0.02 &$-$1.25$\pm$0.01 &$-$6.28$\pm$0.07 &!--/!0/!0/!0     \\
RXC J1532+3021   & 1.22$\pm$0.02  & 1.30$\pm$0.02   &...              & 0.91$\pm$0.08  &26/14/!--/!6 &   0.28$\pm$0.04 &$-$0.27$\pm$0.01 &$-$2.97$\pm$1.28 &$-$8.66$\pm$0.25 &28/!0/!0/!0      \\[1.1mm]
RXC J2003--2323  & 0.97$\pm$0.08  & 0.96$\pm$0.01   & 0.95$\pm$0.02   &...             &!4/!5/!3/!-- &...              &$-$1.22$\pm$0.02 &$-$0.73$\pm$0.01 &$-$6.79$\pm$0.16 &!--/29/!0/!0     \\
RX J1347--1145   & 1.65$\pm$0.05  & 1.63$\pm$0.01   & 1.04$\pm$0.02   & 1.27$\pm$0.06  &!4/14/!3/!6  &...              &$-$0.40$\pm$0.01 &$-$1.78$\pm$0.01 &$-$6.80$\pm$0.08 &!--/!0/!0/!0     \\
RX J1720+2638    & 0.87$\pm$0.03  & 0.96$\pm$0.01   & 0.77$\pm$0.03   &...             &12/!7/!3/!-- &   0.33$\pm$0.03 &$-$0.33$\pm$0.01 &$-$2.99$\pm$0.45 &$-$7.81$\pm$0.14 &28/!0/!0/!0      \\
RX J2129+0005    & 1.07$\pm$0.13  & 1.00$\pm$0.02   & 0.64$\pm$0.06   & 0.82$\pm$0.07  &!1/14/!3/!6  &   0.42$\pm$0.04 &$-$0.40$\pm$0.01 &$-$2.43$\pm$0.11 &$-$7.11$\pm$0.05 &28/!0/!0/!0      \\
S780             & 1.19$\pm$0.09  & 0.94$\pm$0.01   & 0.89$\pm$0.03   &...             &!4/!2/!3/!-- &...              &$-$0.36$\pm$0.01 &$-$2.42$\pm$0.11 &$-$7.49$\pm$0.17 &!--/!0/!0/!0     \\[1.1mm]
Toothbrush       & 1.00$\pm$0.09  & 0.96$\pm$0.09   & 1.03$\pm$0.02   &...             &27/!0/!--/!--&...              &$-$1.03$\pm$0.01 &$-$1.18$\pm$0.01 &$-$6.38$\pm$0.03 &!--/!0/!0/!0     \\ 
Z3146            & 1.29$\pm$0.04  & 1.28$\pm$0.02   & 0.81$\pm$0.03   & 0.91$\pm$0.06  &!8/14/!--/!6 &   0.39$\pm$0.02 &$-$0.34$\pm$0.01 &$-$2.31$\pm$0.03 &$-$8.03$\pm$0.11 &!0/!0/!0/!0      \\ 
Z5247            & 0.80$\pm$0.12  & 0.63$\pm$0.03   & 0.77$\pm$0.03   & 0.85$\pm$0.11  &!1/14/!3/!6  &$-$0.09$\pm$0.05 &...              &...              &...              &28/!--/!--/!--   \\
ZwCl0008+5215    &-0.30$\pm$0.12  &-0.34$\pm$0.12   & 0.53$\pm$0.05   &...             &27/!0/!3/!-- &...              &...              &...              &...              &!--/!--/!--/!--  \\
ZwCl2341+0000    & 0.39$\pm$0.09  & 0.35$\pm$0.09   & 0.71$\pm$0.04   &...             &25/!0/!3/!-- &$-$0.56$\pm$0.08 &$-$1.08$\pm$0.04 &$-$0.83$\pm$0.01 &$-$5.76$\pm$0.19 &28/!0/!0/!0      \\[1.1mm] 
\hline
\end{tabular}
\end{center}
}
{Notes. Columns: (1) cluster name; (2)\textendash(5) mass proxies of cluster, $L_{\rm X}$ and $L_{500}$ in $10^{44}\rm erg~s^{-1}$, and cluster masses $M_{\rm SZ,~500}$ and $M_{500}$ in $10^{14}M_{\odot}$. The uncertainty with mark * for mass proxy is not available from the reference, and 30\% of the total luminosity is taken here; (6) references of mass proxies or mass; (7)\textendash(10) optical and X-ray dynamical parameters, while $\log \omega$ and $\log(P_{3}/P_{0})$ are calculated in 500 kpc; (11) references for dynamical parameters. Clusters hosting both radio halo and relic are not listed twice.\\
References: 0\textemdash this paper by the authors; 1\textemdash \citet{EEB+98}; 2\textemdash \citet{PAPP+11}; 3\textemdash \citet{PAA+15}; 4\textemdash \citet{BSGC+04}; 5\textemdash \citet{CEB+13}; 6\textemdash \citet{WH15}; 7\textemdash \citet{ZLC+15}; 8\textemdash \citet{BVH+00}; 9\textemdash \citet{PBB+04}; 10\textemdash \citet{VBR+11}; 11\textemdash \citet{DFJ+99}; 12\textemdash \citet{EVB+96}; 13\textemdash \citet{RB+02}; 14\textemdash \citet{MAE+10}; 15\textemdash \citet{FGG+12}; 16\textemdash \citet{EMT+02}; 17\textemdash \citet{KEMT+07}; 18\textemdash \citet{EBDM+07}; 19\textemdash \citet{BDR+11}; 20\textemdash \citet{BIB+14}; 21\textemdash \citet{MHS+12}; 22\textemdash \citet{BBVV+12}; 23\textemdash \citet{BBVV+12}; 24\textemdash \citet{BGFG+09}; 25\textemdash \citet{DVB+14}; 26\textemdash \citet{EEMB+10}; 27\textemdash \citet{VAB+99}; 28\textemdash \citet{WH+13}; 29\textemdash \citet{CEGB+10}.}
\label{tab2}
\end{table*}

Throughout this paper, we assume a $\Lambda$CDM cosmology, taking
$H_0=$100 $h$ ${\rm km~s}^{-1}$ ${\rm Mpc}^{-1}$, with $h=0.7$,
$\Omega_m=0.3$, and $\Omega_{\Lambda}=0.7$. Derived parameters in the
literature have been scaled to this cosmology.

\begin{figure*}
\centering\includegraphics[bb= 49 49 999 1438,angle=-90,clip,width=0.85\textwidth]{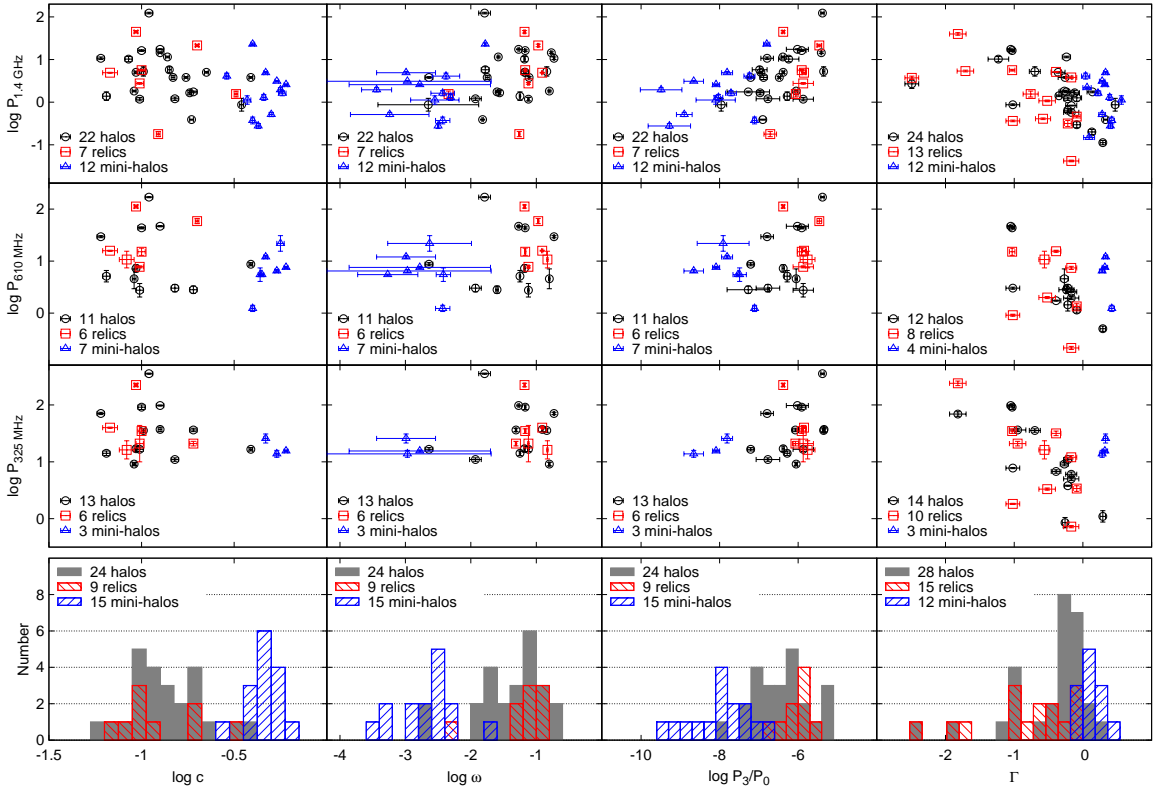}
\caption{The radio power at three frequencies of radio halos ({\it
    circles}), relics ({\it squares}), and mini-halos ({\it
    triangles}) are plotted dynamic parameters of galaxy clusters
  ({\it panels in the upper 3 rows}), and the distributions dynamic
  parameters are shown in the {\it bottom panels}.}
    \label{fig1}
\end{figure*}

\begin{figure*}
\centering\includegraphics[angle=-90,width=0.90\textwidth]{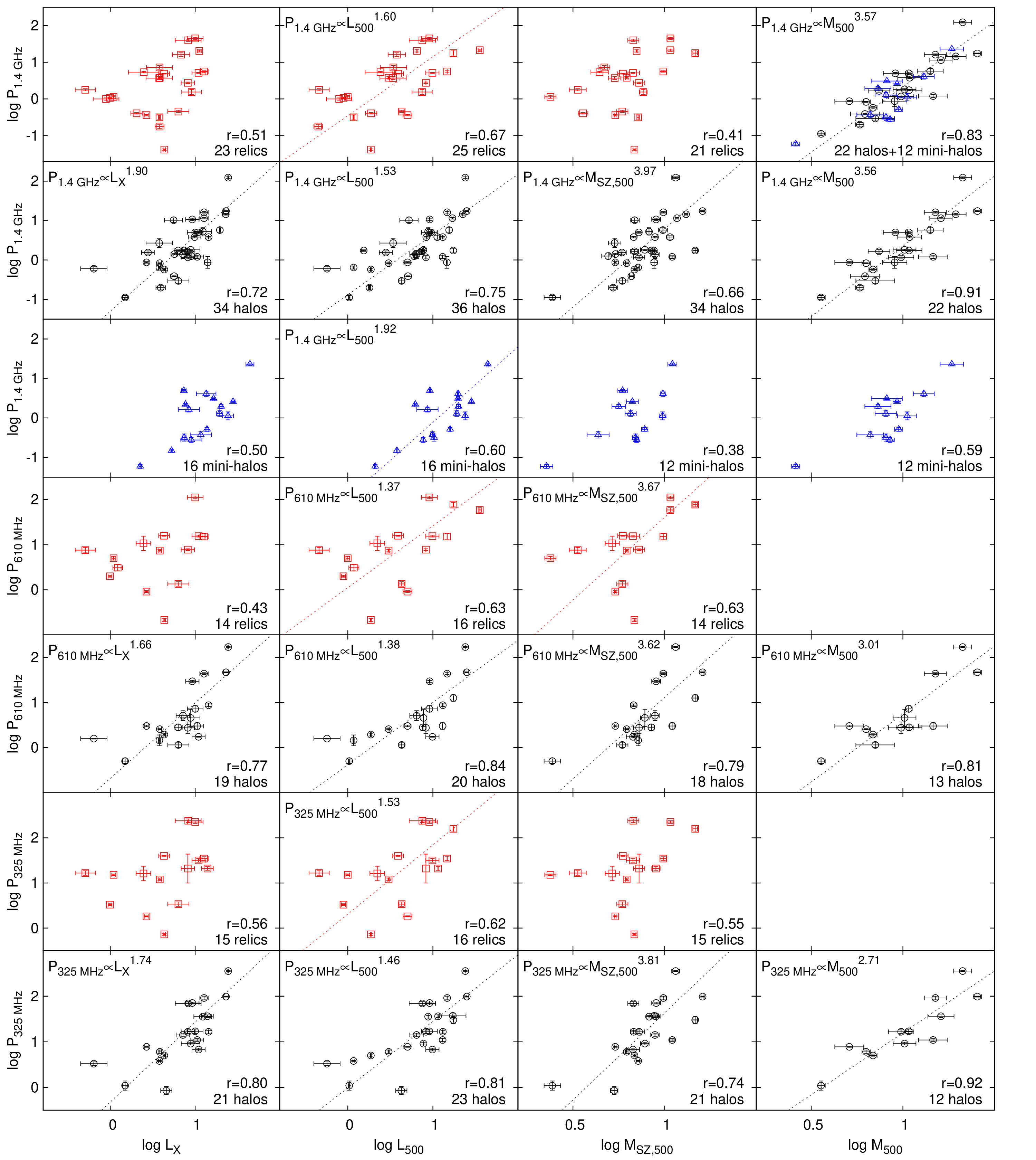}
\caption{The scaling relations for radio power of radio relics and
  halos with cluster masses or mass proxies at three frequencies.
  Plots are omitted if there are only few ($<10$) data points, e.g.,
  those for mini-halos at two lower frequencies and those for relics
  against $M_{500}$. Dotted lines are the best fits, carried out only
  if the Spearman rank-order correlation coefficient $r\ga 0.6$. The
  radio powers $P_{\rm 1.4~GHz}$ of 22 halos and 12 mini-halos are
  plotted together in the right panel of the first row (originally for
  relics) to show their consistency with the scaling relations.}
  \label{fig2}
\end{figure*}

\section{Three sets of data for galaxy clusters}

In this section, we collect and rescale the values of radio flux and
power in Table 1 and cluster mass and the cluster dynamical state for
75 galaxy clusters in Table 2 for further analyses.

\subsection{Radio Power of Radio Halos, Relics, and Mini-halos}

A large number of radio halos, relics, and mini-halos have been
discovered and measured in recent decades through observations with
VLA \citep[e.g.,][]{GF+00,VBR+11}, GMRT
\citep[e.g.,][]{VGBC+07,KVG+15}, WSRT \citep[e.g.,][]{VRB+10,TAL+15}
and also ATCA \citep[e.g.,][]{SBF+14,SMB+15}. We have checked the
radio images of radio halos, relics, and mini-halos in the literature
and collected in Table~1 the radio flux $S_\nu$ at frequencies within
a few per cent around 1.4~GHz, 610~MHz, and 325~MHz; we have
interpolated the flux at an intermediate frequency if measurements are
available at higher and lower frequencies. To establish the reliable
scaling relations, we include only the very firm detection of diffuse
radio emission in galaxy clusters, and omit the questionable
detections or flux estimates due to problematic point-source
subtraction. We then calculate the radio power via
\begin{equation}
P_{\nu}=4\pi D_{\rm L}^{2}\times S_{\nu}\times (1+z)^{1-a},
\label{power}
\end{equation}
where $D_{\rm L}=(1+z) c/H_0 \int_{0}^{z}\frac{dz'}
{\sqrt{\Omega_{m}(1+z')^3+\Omega_{\Lambda}}}$ is the luminosity
distance of a cluster at a redshift of $z$, $S_{\nu}$ is the radio
flux at frequency $\nu$, $(1+z)^{(1-a)}$ is the {\it k}-correction
term as done by \citet{CEB+13}, and $a$ is the spectral index of
diffuse radio sources, which is assumed to be 1.3 in general.

\subsection{Mass Proxies and Mass Estimates for Galaxy Clusters}

The total X-ray luminosities, $L_{\rm X}$, of galaxy clusters and the
X-ray luminosities, $L_{500}$, within $R_{500}$ are most often used as
mass proxies for galaxy clusters.  Here, $R_{500}$ is the radius of a
galaxy cluster within which the matter density of a cluster is 500
times of the critical density of the universe. In Table 2, we collect
these two X-ray measurements for galaxy clusters with diffuse radio
emission. The total X-ray luminosities of galaxy clusters, $L_{\rm
  X}$, were derived from observations in the 0.1-2.4 keV band and
taken from catalogs based on the {\it ROSAT} All-Sky Survey data
\citep[e.g.,][]{EVB+96,EEB+98,EEAC+00,BVH+00,BSGC+04}.  The collected
X-ray luminosities within $R_{500}$ of the clusters, $L_{500}$, are
the values updated by using the new measurements from deep {\it
  Chandra} or {\it XMM-Newton} images from \citet{MAE+10},
\citet{CEB+13}, and \citet{ZLC+15}.

Masses of galaxy clusters can be estimated from the SZ measurements of
the integrated Compton parameter $Y_{\rm SZ,~500}$ within $R_{500}$ via
\begin{equation}
Y_{\rm SZ,500}=d_{\rm
  A}^{2}(z)Y=Q(\Omega_{m}(1+z)^3+\Omega_{\Lambda})^{1/3}M_{\rm
  SZ,500}^{ \kappa},
\label{y500}
\end{equation}
where $d_{\rm A}(z)$ is the angular diameter distance to clusters, $Y$
is the integrated Compton parameter, and $M_{\rm SZ,~500}$ is the mass
within $R_{500}$ estimated from the SZ effect, $\log Q=-0.19$, and
$\kappa=1.79$ \citep{PAA+14}. Note that the $Y_{\rm SZ,~500}$ and
$M_{\rm SZ,~500}$ are scaled by a power index $\kappa=1.79$. The
largest SZ-selected catalog to date is the all-sky Planck catalog of
galaxy clusters, which contains 1653 clusters with redshifts up to
$z\sim1$ \citep{PAA+15}. In this paper we take the mass estimates
$M_{\rm SZ,~500}$ directly from \citet{PAA+15} for galaxy clusters.

In the literature, cluster mass $M_{500}$ has often been derived by
using three X-ray proxies: average temperature $T_{\rm X}$, gas mass
$M_{\rm gas}$, and $Y_{\rm X}=T_{\rm X}\times M_{\rm gas}$
\citep[e.g.,][]{VBE+09,ZLC+15}. Another mass estimate used in this
paper as one of four independent mass proxies in Table~2 is the
cluster mass derived from the observed gas mass. We take $M_{500}$
from \citet{VBE+09} and \citet{MAE+10}, obtained from the high-quality
X-ray images and spectra of {\it Chandra} and {\it XMM} data. The
systematic offset between the mass values in these two catalogs has
been corrected according to \citet{WH15}.

\subsection{Dynamical Parameters of Galaxy Clusters}

\citet{WH+13} developed a method to quantify dynamical states of
galaxy clusters from optical photometric data. They smoothed the
brightness distribution of member galaxies using a Gaussian kernel
with a weight of optical luminosity, and then defined a dynamical
parameter $\Gamma$ from the asymmetry, the normalized model-fitting
residual, and the ridge-flatness of the smoothed optical image. They
obtained $\Gamma$ values for 98 clusters with qualitatively known
``relaxed'' or ``unrelaxed'' dynamical states, and then also for 2092
rich clusters of $M_{200} \ge 3.15 \times 10^{14} M_{\odot}$ in the
cluster catalog of \citet{WHL+12}. We quoted $\Gamma$ in Table~2
from \citet{WH+13} for 58 galaxy clusters with detected radio halos,
relics, and mini-halos, and also calculated $\Gamma$ values for the
remaining 23 galaxy clusters that are not included in \citet{WH+13}.

\begin{table*}[t]
\setlength{\tabcolsep}{1.5mm}
{\footnotesize
\begin{center}
\caption{The Scaling Relation for Radio Power of Relics, Radio Halos,
  and Mini-halos in Galaxy Clusters Together with the Intrinsic Data
  Scatter $\rm \sigma^2/dof$ and the Fitting $\rm \chi^2/dof$.}
\begin{tabular}{llclllcc}
\hline
\hline
\mc{1}{l}{Parameters} &\mc{1}{c}{No.} &\mc{1}{c}{Type} &\mc{1}{c}{$r$} &\mc{1}{c}{$p$} &\mc{1}{l}{The Best Fitted Relations} &\mc{1}{c}{$\rm \sigma^2/dof$} &\mc{1}{c}{$\rm \chi^2/dof$}\\
\hline
$P_{\rm1.4~GHz}$--$L_{500}$            &25  &relic     &0.67 &0.00 &$\log P_{\rm1.4~GHz}$=$(1.60\pm0.17)\log L_{500}$--$(0.46\pm0.24)$       &0.474 &0.965\\
$P_{\rm610~MHz}$--$L_{500}$            &16  &relic     &0.63 &0.01 &$\log P_{\rm610~MHz}$=$(1.37\pm0.13)\log L_{500}$+$(0.05\pm0.22)$       &0.478 &0.983\\
$P_{\rm610~MHz}$--$M_{\rm SZ,~500}$      &14  &relic     &0.63 &0.02 &$\log P_{\rm610~MHz}$=$(3.67\pm0.20)\log M_{\rm SZ,~500}$--$(2.06\pm0.24)$ &0.615 &0.988\\
$P_{\rm325~MHz}$--$L_{500}$            &16  &relic     &0.62 &0.01 &$\log P_{\rm325~MHz}$=$(1.53\pm0.15)\log L_{500}$+$(0.31\pm0.23)$       &0.530 &0.985\\
\hline
$P_{\rm1.4~GHz}$--$L_{\rm X}$           &34  &halo      &0.72 &0.00 &$\log P_{\rm1.4~GHz}$=$(1.90\pm0.14)\log L_{\rm X}$--$(1.29\pm0.18)$     &0.250 &0.868\\
$P_{\rm1.4~GHz}$--$L_{500}$            &36  &halo      &0.75 &0.00 &$\log P_{\rm1.4~GHz}$=$(1.53\pm0.13)\log L_{500}$--$(0.88\pm0.18)$       &0.226 &0.949\\
$P_{\rm1.4~GHz}$--$M_{\rm SZ,~500}$      &34  &halo      &0.66 &0.00 &$\log P_{\rm1.4~GHz}$=$(3.97\pm0.18)\log M_{\rm SZ,~500}$--$(3.18\pm0.20)$ &0.263 &0.969\\
$P_{\rm1.4~GHz}$--$M_{500}$            &22  &halo      &0.91 &0.00 &$\log P_{\rm1.4~GHz}$=$(3.56\pm0.12)\log M_{500}$--$(3.21\pm0.15)$       &0.139 &0.746\\[1mm]
$P_{\rm610~MHz}$--$L_{\rm X}$           &19  &halo      &0.77 &0.00 &$\log P_{\rm610~MHz}$=$(1.66\pm0.13)\log L_{\rm X}$--$(0.67\pm0.17)$     &0.246 &0.904\\
$P_{\rm610~MHz}$--$L_{500}$            &20  &halo      &0.84 &0.00 &$\log P_{\rm610~MHz}$=$(1.38\pm0.12)\log L_{500}$--$(0.38\pm0.16)$       &0.206 &0.965\\
$P_{\rm610~MHz}$--$M_{\rm SZ,~500}$      &18  &halo      &0.79 &0.00 &$\log P_{\rm610~MHz}$=$(3.62\pm0.16)\log M_{\rm SZ,~500}$--$(2.50\pm0.17)$ &0.241 &0.983\\
$P_{\rm610~MHz}$--$M_{500}$            &13  &halo      &0.81 &0.00 &$\log P_{\rm610~MHz}$=$(3.01\pm0.11)\log M_{500}$--$(2.27\pm0.14)$       &0.165 &0.823\\[1mm]
$P_{\rm325~MHz}$--$L_{\rm X}$           &21  &halo      &0.80 &0.00 &$\log P_{\rm325~MHz}$=$(1.74\pm0.12)\log L_{\rm X}$--$(0.31\pm0.15)$     &0.229 &0.873\\
$P_{\rm325~MHz}$--$L_{500}$            &23  &halo      &0.81 &0.00 &$\log P_{\rm325~MHz}$=$(1.46\pm0.11)\log L_{500}$--$(0.01\pm0.15)$       &0.198 &0.926\\
$P_{\rm325~MHz}$--$M_{\rm SZ,~500}$      &21  &halo      &0.74 &0.00 &$\log P_{\rm325~MHz}$=$(3.81\pm0.16)\log M_{\rm SZ,~500}$--$(2.18\pm0.17)$ &0.240 &0.963\\
$P_{\rm325~MHz}$--$M_{500}$            &12  &halo      &0.92 &0.00 &$\log P_{\rm325~MHz}$=$(2.71\pm0.07)\log M_{500}$--$(1.52\pm0.09)$       &0.111 &0.789\\[1mm]
\hline
$P_{\rm 1.4~GHz}-L_{500}$            &16  &mini-halo &0.60 &0.01 &$\log P_{\rm 1.4~GHz}$=$(1.92\pm 0.10)\log L_{500}$--$(2.03\pm 0.15)$       &0.211 &0.989\\
\hline
\end{tabular}
\end{center}
\label{tab3}
}
\end{table*}

Dynamical parameters have also been derived quantitatively from X-ray
images of clusters by previous authors, including the concentration
parameter $c$ \citep[e.g.,][]{SRT+08}, the centroid shift $\omega$
\citep[e.g.,][]{PFB+06}, and the power ratio $P_3/P_0$
\citep[e.g.,][]{BT+95,BPA+10,WBSA+13}. The concentration parameter $c$
is defined as the ratio of the peak to the ambient surface brightness
as
\begin{equation}
c=\frac{S(R<\rm 100kpc)}{S(R<\rm 500kpc)}.
\label{c}
\end{equation} 
The centroid shift $\omega$ is defined as the standard deviation of
the projected separation between the X-ray peak and the centroid in
units of $R_{\rm ap}=\rm 500kpc$, which is computed in a series of
circular apertures centered on the X-ray peak from $R_{\rm ap}$ to
$0.05R_{\rm ap}$ in steps of $0.05R_{\rm ap}$, thus
\begin{equation}
\omega=[\frac{1}{N-1}\sum(\Delta_{i}-\langle \Delta
  \rangle)^2]^{1/2}\times \frac{1}{R_{\rm ap}}.
\label{w}
\end{equation}
Here $\Delta_{i}$ is the distance between the X-ray peak and the
centroid of the $i$th aperture \citep{PFB+06}. \citet{BT+95} defined
the power ratios as dimensionless morphological parameters from the
two-dimensional multipole expansion of the projected gravitational
potential of clusters inside $R_{\rm ap}$. The moments, $P_{m}$, are
defined as follows:
\begin{equation}
P_{0}=[a_{0}\ln (R_{\rm ap})]^2,
\label{p0}
\end{equation}
\begin{equation}
P_{m}=\frac{1}{2m^{2}R_{\rm ap}^{2m}}(a_{m}^2+b_{m}^2).
\label{pm}
\end{equation}
The moments $a_{m}$ and $b_{m}$ are calculated using
\begin{equation}
a_{m}(R)=\int_{R'\le R_{\rm ap}}S(x')(R')^{m}\cos(m\phi')d^2x',
\label{am}
\end{equation}
and
\begin{equation}
b_{m}(R)=\int_{R'\le R_{\rm ap}}S(x')(R')^{m}\sin(m\phi')d^2x',
\label{bm}
\end{equation}
where $S(x)$ is the X-ray surface brightness of the pixel labeled
$x$. $P_{3}/P_{0}$ is the power ratio, which was found to be related
to substructures \citep[e.g.,][]{BPA+10,CEGB+10}. We therefore also
take $P_{3}/P_{0}$ as another dynamical parameter of clusters.

The dynamical parameters in Table 2 are taken directly from the
literature for the galaxy clusters that have diffuse radio emission.
For 49 clusters, we derive the concentration parameters, $c$, the
centroid shifts, $\omega$, and the power ratios, $P_{3}/P_{0}$, from
the {\it Chandra} 0.5-5 keV band X-ray
images\footnote{http://cda.harvard.edu/chaser/} by using Equations
(3)\textendash(8).  We take our newly derived dynamical parameters if
they are different from the values given in the literature.

\section{The scaling relations for radio power and 
the fundamental plane in the 3D parameter space}

The data distribution of the three sets of parameters is shown in
Figure~\ref{fig1}. In general, the values of radio power for the three
types of diffuse emission in galaxy clusters are in the same range of
magnitude. 

The ranges of dynamical parameters for clusters with radio halos and
mini-halos in Figure~\ref{fig1} are consistent with those of
\citet[][Figure~1]{CEGB+10}. In particular, we found that galaxy
clusters with mini-halos have very large $c$ and $\Gamma$ ($\log c \ga
-0.5$, $\Gamma \ga -0.2$) and a small $\omega$ and $P_3/P_0$ ($\log
\omega \la -2$, $\log(P_3/P_0) \la -7$), indicating the relaxed state
of these clusters. Clusters with relics and halos share quite similar
dynamical properties. The $\Gamma$ distributions show clusters with
relics to be more disturbed than clusters with radio halos, which is
probably related to the fact that radio relics are likely found in
clusters characterized by mergers happening almost on the plane of the
sky.

Clusters with relics have a slightly wider range of X-ray luminosity
and hence a larger range of masses than those with halos (see Figure
2), while clusters with radio mini-halos have a slightly smaller range
of higher X-ray luminosity.

In the following we discuss the scaling relations in the
two-dimensional data distributions for the radio power, and then try
to find the fundamental plane in three-dimensional parameter spaces.
The Bivariate Correlated Errors and intrinsic Scatter (BCES) method
has previously been used in similar analyses
\citep[e.g.,][]{BCD+09,CEB+13}. We develop the BCES-Reduced Major Axis
(BCES-RMA) method for the three-dimensional data fitting (see the
appendix for details), and use the BCES-RMA in the following to get
the regression parameters for 2D and 3D fittings. The unified
deviations $\rm\sigma^2/dof$ for the intrinsic scatter (see
Equation~(\ref{eqA11}) in the appendix, not including the contribution
from measurement uncertainties) as well as the fitting $\rm\chi^2/dof$
(see Equation~(\ref{eqA13}) in the appendix) are calculated
accordingly. In addition, we use the Spearman rank-order correlation
coefficient, $r$, to assess data correlations and the probability of
the null hypothesis $p$ to indicate the reliability of correlations
\citep[see][p. 634]{ptvf92}. For 3D fittings, we first compute
$\hat{z}_i$ from variables $x_i$ and $y_i$ based on the 3D best
fitting relations, and then calculate the coefficient $r$ from
$\hat{z}_i$ and variables $z_i$.

\begin{table*}[t]
\setlength{\tabcolsep}{0.5mm}
{\footnotesize
\begin{center}
\caption{Searching for a Fundamental Plane in 3D Parameter Space 
    by Involving Dynamical Parameters and Comparing the Intrinsic Data
    Scatter $\rm\sigma^2/dof$ and the Fitting $\rm\chi^2/dof$.}
\begin{tabular}{lcccclcc}
\hline
\hline
\mc{1}{l}{Parameters}     &\mc{1}{c}{No.}     &\mc{1}{c}{Type}   &\mc{1}{c}{$r$}   &\mc{1}{c}{$p$}   &\mc{1}{l}{The best-fitted relation}     &\mc{1}{c}{$\rm \sigma^2/dof$}     &\mc{1}{c}{$\rm \chi^2/dof$}\\
\hline
$P_{\rm 1.4~GHz}$--$L_{500}$            &13  &halo      &0.52 &0.07 &$\log P_{\rm 1.4~GHz}$=$(2.56\pm 0.11)\log L_{500}$--$(2.03\pm 0.15)$                              &0.309 &0.903\\
$P_{\rm 1.4~GHz}$--$L_{500}$--$\Gamma$  &13  &halo      &0.54 &0.06 &$\log P_{\rm 1.4~GHz}$=$(1.03\pm 0.05)\log L_{500}$--$(0.87\pm 0.17)\Gamma$--$(0.83\pm 0.06)$      &0.050 &1.002\\
$P_{\rm 1.4~GHz}$--$L_{500}$--$c$       &13  &halo      &0.64 &0.02 &$\log P_{\rm 1.4~GHz}$=$(1.86\pm 0.08)\log L_{500}$--$(2.39\pm 0.16)\log c$--$(3.30\pm 0.10)$      &0.154 &0.989\\
$P_{\rm 1.4~GHz}$--$L_{500}$--$\omega$  &13  &halo      &0.63 &0.02 &$\log P_{\rm 1.4~GHz}$=$(2.05\pm 0.06)\log L_{500}$+$(1.00\pm 0.07)\log \omega$--$(0.07\pm 0.07)$  &0.143 &0.854\\
$P_{\rm 1.4~GHz}$--$L_{500}$--$\frac{P_{3}}{P_{0}}$&13&halo&0.68&0.00 &$\log P_{\rm 1.4~GHz}$=$(1.57\pm 0.08)\log L_{500}$+$(0.61\pm 0.01)\log \frac{P_{3}}{P_{0}}$+$(2.99\pm 0.09)$ &0.160 &0.864\\[1mm]
$P_{\rm 1.4~GHz}$--$L_{500}$            &24  &halo      &0.70 &0.00 &$\log P_{\rm 1.4~GHz}$=$(1.56\pm 0.13)\log L_{500}$--$(0.94\pm 0.16)$                              &0.212 &0.955\\
$P_{\rm 1.4~GHz}$--$L_{500}$--$\Gamma$  &24  &halo      &0.75 &0.00 &$\log P_{\rm 1.4~GHz}$=$(1.07\pm 0.13)\log L_{500}$--$(0.52\pm 0.29)\Gamma$--$(0.78\pm 0.13)$      &0.130 &0.955\\
$P_{\rm 1.4~GHz}$--$M_{\rm SZ,~500}$          &24  &halo      &0.64 &0.00 &$\log P_{\rm 1.4~GHz}$=$(3.69\pm 0.16)\log M_{\rm SZ,~500}$--$(2.94\pm 0.16)$                            &0.209 &0.958\\
$P_{\rm 1.4~GHz}$--$M_{\rm SZ,~500}$--$\Gamma$&24  &halo      &0.68 &0.00 &$\log P_{\rm 1.4~GHz}$=$(2.68\pm 0.13)\log M_{\rm SZ,~500}$--$(0.55\pm 0.28)\Gamma$--$(2.28\pm 0.12)$    &0.113 &0.946\\[1mm]
$P_{\rm 1.4~GHz}$--$M_{500}$            &17  &halo      &0.88 &0.00 &$\log P_{\rm 1.4~GHz}$=$(3.19\pm 0.10)\log L_{500}$--$(2.87\pm 0.12)$                              &0.125 &0.734\\
$P_{\rm 1.4~GHz}$--$M_{500}$--$\Gamma$  &17  &halo      &0.90 &0.00 &$\log P_{\rm 1.4~GHz}$=$(1.75\pm 0.12)\log L_{500}$--$(0.43\pm 0.11)\Gamma$--$(1.61\pm 0.10)$      &0.090 &0.871\\
\hline
$P_{\rm 1.4~GHz}$--$L_{500}$            &13  &relic     &0.62 &0.02 &$\log P_{\rm 1.4~GHz}$=$(2.18\pm 0.16)\log L_{500}$--$(1.05\pm 0.22)$                              &0.626 &0.947\\
$P_{\rm 1.4~GHz}$--$L_{500}$--$\Gamma$  &13  &relic     &0.71 &0.01 &$\log P_{\rm 1.4~GHz}$=$(0.88\pm 0.25)\log L_{500}$--$(0.45\pm 0.13)\Gamma$--$(0.71\pm 0.16)$      &0.353 &0.984\\[1mm]
\hline
$P_{\rm 1.4~GHz}$--$L_{500}$            &12  &mini-halo &0.48 &0.11 &$\log P_{\rm 1.4~GHz}$=$(2.31\pm 0.08)\log L_{500}$--$(2.58\pm 0.12)$                              &0.230 &0.960\\
$P_{\rm 1.4~GHz}$--$L_{500}$--$\frac{P_{3}}{P_{0}}$&12&mini-halo&0.66 &0.02 &$\log P_{\rm 1.4~GHz}$=$(1.37\pm 0.08)\log L_{500}$+$(0.32\pm 0.01)\log \frac{P_{3}}{P_{0}}$+$(1.14\pm 0.10)$&0.172 &0.964\\[1mm]
\hline
\end{tabular}
\end{center}
\label{tab4}
}
\end{table*}

\begin{figure*}
  \centering
  \includegraphics[angle=-90,width=0.98\textwidth]{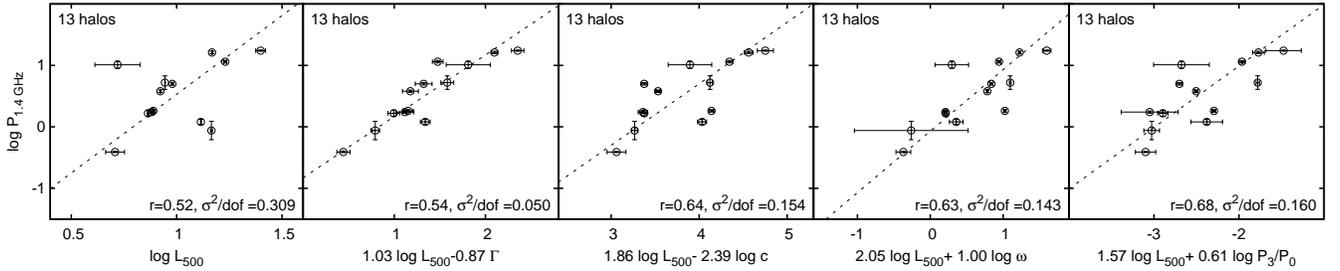}
  \caption{Comparison of the effectiveness by involving different
    dynamical parameters to reduce the data scatter.}
  \label{fig3}
\end{figure*}

\subsection{The Scaling Relations between Radio Power and Cluster Mass}

The scaling relation between radio power of radio halos and mass
proxies of galaxy clusters has been studied by many authors, e.g.,
\citet{LHB+00}, \citet{BCD+09}, \citet{Ba+12}, and \citet{CEB+13}. This
relation can be written as
\begin{equation}
\log P_{\rm 1.4~GHz}=\alpha \log M+C,
\label{scale2d}
\end{equation}
where $C$ is the normalization factor, $M$ is the mass parameter of
clusters, and $\alpha$ is the index. \citet{BCD+09} took the X-ray
luminosity $L_{\rm X}$ as the mass proxy for clusters and obtained
$\alpha_{L_{\rm X}}=2.06\pm0.20$ for 22 halos and two
mini-halos. \citet{CEB+13} obtained $L_{500}$ from the {\it Chandra}
images for 25 clusters with halos and found
$\alpha_{L_{500}}=2.11\pm0.20$. By using the SZ parameter $Y_{500}$ as
a mass proxy, they obtained $\alpha_{Y_{500}}=2.02\pm0.28$ for these
clusters, and then found $\alpha_{M_{500}}=3.70\pm0.56$ for $M_{500}$.
Since cluster mass $M_{500}$ is related to $L_{500}$ by
$ L_{500} \propto M_{500}^{1.64}$ \citep{PAPP+11},
it is understandable that $\alpha_{M_{500}} = \alpha_{L_{500}} \times
1.64$.

By using the radio power values of halos, relics, and mini-halos at
the three frequencies in Table~1 and cluster masses or proxies in
Table~2, we check the scaling relations between the radio power and
cluster masses for galaxy clusters. The power values of a pair of
relics detected from one cluster are added for the following
discussions. Results are shown in Figure~\ref{fig2} and listed in
Table~3.

First of all, let us look at different types of radio emission. The
power of radio halos at any frequency is clearly correlated with the
kinds of cluster masses or mass proxies. They show the strongest
correlations and much less intrinsic data scattered around the
best-fit correlations. For the relics and mini-halos, the radio power
is found to be only marginally correlated with $L_{500}$ (and also
with $M_{\rm SZ,~500}$ for relics at 610~MHz), and the correlations
are less strong and also the points are clearly more scattered around
the best-fit correlations, as shown by the $\rm\sigma^2/dof$ in Table
3. The radio power of mini-halos at 1.4~GHz, if plotted against
cluster mass, is consistent with the result in \citet{GMV+14}, but we
find a marginal correlation between $P_{\rm 1.4~GHz}$ and $L_{500}$ or
$M_{500}$ with a Spearman rank-order correlation coefficient $r=0.6$
or 0.59. We also noticed that the radio power $P_{\rm 1.4~GHz}$ of
halos and mini-halos at 1.4~GHz can be scaled together very well with
$M_{500}$, as shown in the right panel of the first row in Fig.2 for
the 22 halos together with 12 mini-halos.
  
Second, which of the mass estimates or mass proxies is good for the
scaling relations? For relics and mini-halos, $L_{500}$ seems to be
the best, because not only are more data available for the host
clusters but also the other masses or proxies do not show significant
correlation. For radio halos, the $M_{500}$ estimated from gas mass is
the best for the scaling relations with radio power at any frequency,
though fewer data are available for host clusters and thus we cannot
exclude that the small size of the sample can affect the strength of
the correlation. Among the other three mass proxies, $L_{500}$ shows a
slightly better correlation with the halo radio power than $L_{\rm X}$
and $M_{\rm SZ,~500}$, as indicated by a slightly larger Spearman
rank-order correlation coefficient $r$ and a smaller deviation $\rm
\sigma^2/dof$ as listed in Table~3. Therefore $L_{500}$ is a common
mass proxy for galaxy clusters which can be scaled with the radio
power of all three types of diffuse radio emission.

We noticed that at any of these three frequencies, the scaling indices
$\alpha_{L_{500}}$ between the radio power and proxy $L_{500}$ are
almost the same for the relics and radio halos, though relic data are
more scattered around the fitted lines. The scaling index we obtained
for $P_{\rm 1.4~GHz}$ of halos and $L_{\rm X}$ is $\alpha_{L_{\rm
    X}}=1.90\pm0.14$, which is consistent with the previous results
around $\alpha_{L_{\rm X}}=2.06\pm0.2$ in \citet{BCD+09}. Our scaling
indices for the power of radio halos $P_{\rm 1.4~GHz}$ against the SZ
mass and $M_{500}$ are $3.97\pm0.18$ and $3.56\pm0.12$, respectively,
which are consistent with the result $\alpha_{M_{500}}=3.70\pm0.56$
obtained by \citet{CEB+13}. For relics, the scaling index we found for
$P_{\rm 1.4~GHz}$ and $L_{500}$ is $\alpha_{L_{500}}=1.60\pm0.17$,
which is very consistent with the most recent result
$\alpha_{M_{500}}=2.83 \pm0.39$ given by \citet{DVB+14} if we consider
$\alpha_{M_{500}} = \alpha_{L_{500}} \times 1.64$. The scaling indices
$\alpha_{L_{500}}$ are roughly consistent at three frequencies if
considering the uncertainties, while scaling indices
$\alpha_{M_{500}}$ are different at three frequencies for the radio
halos, which may be due to selection effect of the small sample and
needs to be verified further in future.

\begin{figure}
\centering\includegraphics[bb = 48 70 1596 1130,angle=-90,width=0.45\textwidth]{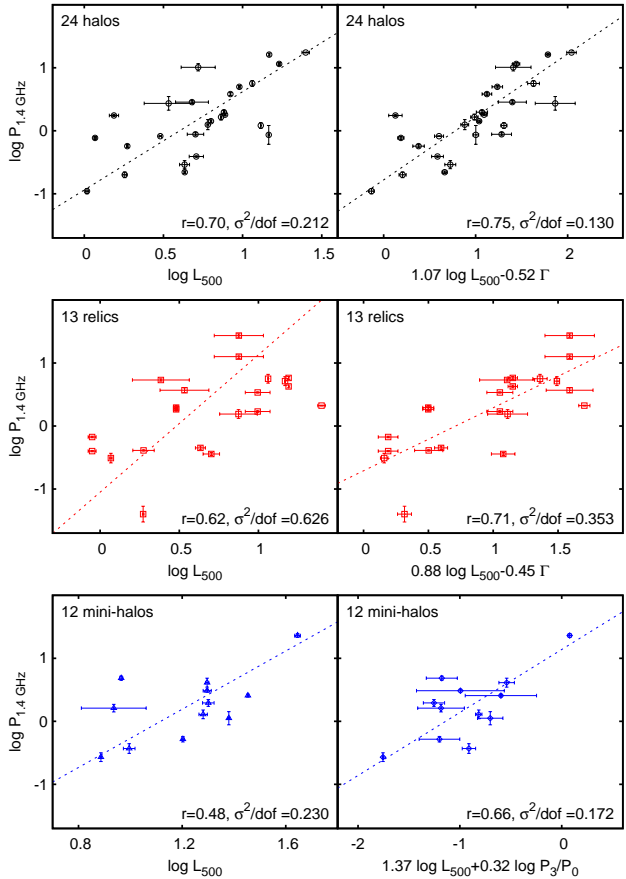}
  \caption{The data scatter is effectively reduced by involving
    dynamical parameters $\Gamma$ for radio halos and relics, and $P_3/P_0$ for
    mini-halos.}
\label{fig4}
\end{figure}

\begin{figure}
  \centering
    \includegraphics[bb = 69 90 362 260,width=0.45\textwidth,angle=0]{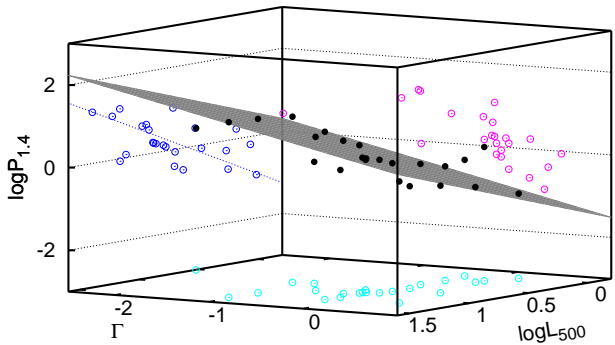}
    \caption{The best-fitted plane for 24 radio halos in 3D parameter
      space: $\log P_{\rm 1.4~GHz} = (1.07\pm 0.13) \log L_{500} -
      (0.52 \pm 0.29)\Gamma -(0.78 \pm 0.13)$. Data are projected onto
      the three planes and shown as open circles.}
    \label{fig5}
\end{figure}

\subsection{Searching for the Fundamental Plane in the 3D Parameter Space}

We search here for the correlation between the radio power $P$ of
halos, relics, and mini-halos with cluster mass $M$ and the dynamical
parameter $D$ in 3D parameter spaces. Based on
Equation~(\ref{scale2d}), the 3D relations in general can be written
as
\begin{equation}
\log P =\alpha \log M+\beta \log D+\gamma, 
\label{3d}
\end{equation}
which is the fundamental plane in the 3D space. The new fitting method
introduced in the appendix can fit data with uncertainties. The data
scatter $\rm\sigma^2/dof$ can be calculated via the offsets from the
plane by considering the data uncertainties (see the appendix).

We search for the fundamental planes separately for radio halos,
mini-halos, and relics. Because there is much less data for radio
power at 610 and 325 MHz, we fit here only the data of $P_{\rm
  1.4~GHz}$. We adopt $L_{500}$ as the main mass proxy, since its
values are available for most galaxy clusters. To make a reasonable
comparison of data scatter among the 2D and 3D correlations, we use
the same cluster subsamples to check whether the inclusion of any
dynamical parameter can reduce the data scatter and improve the fit.
  
First of all, we check which one of the four kinds of dynamical
parameters is most effective. For a subsample of 13 galaxy clusters
with radio halos, all four kinds of dynamical parameters, $\Gamma$,
$\omega$, $c$, and $P_3/P_0$, are available (as listed in Table 2). We
find that involving any one of these dynamical parameters can reduce
the $\sigma^2$/dof of the fitting, as listed in Table 4 and shown in
Figure~\ref{fig3}. Nevertheless the $\Gamma$ can reduce the
$\sigma^2$/dof most significantly from 0.309 to 0.050. In fact, the
dynamical parameter $\Gamma$ is available for a subsample of 24 galaxy
clusters with radio halos, which works effectively as shown in
Figure~\ref{fig4}. The best fitting plane for the 24 radio halos is
shown in Figure~\ref{fig5} in the 3D space of $P_{\rm
  1.4~GHz}-L_{500}-{\Gamma}$. For this subsample of 24 galaxy clusters
with radio halos, the SZ mass estimates $M_{\rm SZ,~500}$ are
available. We found that if we replace $L_{500}$ with $M_{\rm
  SZ,~500}$, $\Gamma$ works similarly well in the 3D fitting, see
Table~4. This is also true for another subsample of 17 galaxy clusters
with $M_{500}$.

We found that the dynamical parameter $\Gamma$ also works well to
reduce the data scatter for also a subsample of 13 galaxy clusters
with relics, as seen in Table 4 and Figure~\ref{fig3}. However, for a
subsample of 12 galaxy clusters with mini-halos, the most effective
dynamical parameter is $P_3/P_0$ which picks up the presence of a cold
front in the X-ray images of cool-core clusters as a signature of gas
sloshing \citep[e.g.,][]{MG08}.

\section{Conclusions and Discussions }

In this paper, we collect the observed fluxes of radio halos, relics,
and mini-halos of galaxy clusters from the literature and calculate
the radio power of these three types of diffuse radio emission at
three frequencies, $P_{\rm 1.4~GHz}$, $P_{\rm 610~MHz}$, and $P_{\rm
  325~MHz}$. We also collect the mass estimates and mass proxies,
$L_{\rm X}$, $L_{500}$, $M_{\rm SZ,~500}$, and $M_{500}$ for these
galaxy clusters, and obtain their dynamical parameters, $\Gamma$, $c$,
$\omega$, and $P_{3}/P_{0}$ from optical and X-ray image data.  The
data show that galaxy clusters with relics, radio halos, and
mini-halos are in different dynamical states described by dynamical
parameters. Radio relics and halos are detected from merging clusters,
and mini-halos from relaxed clusters. By using these data, we studied
the scaling relations for relics, radio halos, and mini-halos and
searched for the fundamental plane in the 3D parameter space.

We conclude from our data that the radio powers of relics, radio
halos, and mini-halos are all correlated with mass proxies
$L_{500}$. The power of radio halos shows the strongest
correlations. For the relics and mini-halos the correlations are less
strong and also the points are clearly more scattered around the
best-fit correlations. For radio halos, the scaling indices between
the radio power and the mass proxies $L_{500}$ and $M_{\rm SZ,~500}$ are
consistent with each other at three frequencies. The powers of radio
halos and mini-halos can be scaled together nicely with the cluster
mass $M_{500}$.

We found that when any of various dynamical parameters is involved,
the data scatter of the scaling relations between the radio power and
mass proxies can be significantly reduced. For radio halos and relics,
the most effective is to include the dynamical parameter $\Gamma$
derived from the optical brightness distribution of cluster member
galaxies. For the mini-halos, the radio power is closely related to
$P_{3}/P_{0}$ derived for the inner X-ray substructures of
globally relaxed clusters.

Evidently the properties of diffuse radio emission in galaxy clusters
are related not only to cluster mass but also to the dynamic states.
First of all, to host diffuse radio emission, a galaxy cluster has to
be massive enough to contain enough intracluster medium for dynamical
stirring either in the central region of relaxed clusters for
mini-halos or on cluster scales of merging clusters for radio halos or
relics. When a massive cluster appears to be very relaxed with a cool
core, a mini-halo could be produced as long as the substructures of
cold fronts in the X-ray image appear \citep{MG08}, indicating that
the turbulence generated by the gas sloshing of the dark-matter cores
in the cluster potential well \citep[e.g.,][]{ZMB+13} is responsible
for re-accelerating the relativistic electrons for diffuse radio
emission.

Merging of galaxy clusters can generate turbulence on a cluster scale,
which can re-accelerate relativistic particles and produce Mpc-size
radio halos \citep[e.g.,][]{BJ14}. The dynamical states of merging
clusters can be imprinted by substructures in the hot gas distribution
seen in X-ray images or by the unrelaxed velocity distribution of
member galaxies or their irregular brightness distributions. Looking
at two proposed theoretical models for cluster halos: (1) the
secondary model in which the relativistic electrons for synchrotron
emission are the secondary products of the inelastic collision of
thermal protons and cosmic-ray protons in clusters, and (2) the
re-acceleration model in which the relativistic electrons are
re-accerlerated by turbulence in the intracluster medium, we found
that our results show the close relation between the dynamic stirring
and radio halos in the format of a fundamental plane, which no doubt
supports the re-acceleration scenario.

The merging of two massive clusters can also induce peripheral shocks
that re-accelerate particles and compress or amplify the magnetic
fields, so that giant radio relics can be produced in the shock region
of the cluster periphery \citep[e.g.,][]{HB+07,KR+13}. The sky
distribution of member galaxy brightness is physically related to the
dynamics of merging clusters, which has influence on the
re-acceleration of particles in the peripheral shock regions and
consequently is related to the radio power of relics as revealed in
this paper.

In summary, in addition to the known scaling relations between the
radio power and X-ray luminosity, we found that the power of radio
halos and relics is correlated with cluster mass proxies and dynamical
parameters in the form of a fundamental plane.

\acknowledgments 

We thank Dr. Tiziana Venturi and also the referee for very careful
reading of the manuscript and very instructive comments that helped us
to improve the paper significantly. The authors are supported by the
Strategic Priority Research Program ``The Emergence of Cosmological
Structures'' of the Chinese Academy of Sciences, Grant
No. XDB09010200, and the National Natural Science Foundation of China
(11103032, 11473034) and the Young Researcher Grant of National
Astronomical Observatories, Chinese Academy of Sciences.

\bibliographystyle{apj}

\begin{appendix}
\section{The 3D linear regression for data with uncertainties}
\label{}
Linear regression analysis is widely used to study the correlation of
two sets of data. Astronomical data sets usually have measurement
uncertainties. The BCES method has been used for astronomical data
analysis \citep[e.g.,][]{BCD+09,CEB+13,ZJC+13} because: (1)
observational data have uncertainties; (2) uncertainties of data sets
can be dependent; (3) regression lines such as the bisector and the
orthogonal regression (OR) can be obtained easily. See \citet{AB+96}
for details.

In this work, the data sets of galaxy clusters in Tables 1 and 2 have
measurement uncertainties, and the level of uncertainties for
different parameters obtained from different observations can be very
different. For example, the uncertainty of $L_{500}$ from
\citet{CEB+13} derived from the {\it Chandra} data is about a
magnitude smaller than those in the MCXC catalog derived from the {\it
  ROSAT} data. The BCES method can be used to fit data in five
approaches: (1) BCES({\it Y\textbar X}), where the deviations of data
to the fitted line are measured vertically; (2) BCES({\it X\textbar
  Y}), where the deviations are measured horizontally; (3) OR, where
the deviations are measured perpendicularly to the fitted line; (4)
RMA, where the deviations are measured both perpendicularly and
horizontally; (5) BCES bisector, which is the bisector of the
BCES({\it Y\textbar X}) and BCES({\it X\textbar Y}) lines. The last
three approaches are usually recommended because both axises are
considered simultaneously. The BCES bisector method has not yet be
developed for 3D fitting. The BCES-RMA method usually gives very
similiar fitting coefficients to the BCES bisector method.

The 2D BCES-RMA method is derived directly from the ordinary
least-square (OLS) method, which ensures that the sum of deviations
between the data points and the fitted line is as small as possible
\citep[e.g.,][]{IFA+90}. The OLS method is only available for data
fitting without considering data uncertainty. If the variables of
interest are denoted by $X_{1i}, X_{2i}$ and the observed data for
them denoted by $Y_{1i}, Y_{2i}$, we have
\begin{equation}
Y_{1i}=X_{1i}\pm \epsilon_{1i}, Y_{2i}=X_{2i}\pm \epsilon_{2i}
\label{eqA1}
\end{equation}
where $\epsilon_{1i},\epsilon_{2i}$ are uncertainties. The linear
regression model is formulized as
\begin{equation}
X_{2i}=\alpha X_{1i}+\beta.
\label{eqA2}
\end{equation}
According to the OLS method, we can obtain the fitting coefficients
$\alpha_1$ and $\beta_1$ for OLS({\it Y\textbar X}) as
\begin{equation}
\begin{aligned}
\centering
\alpha_1&=\frac{C(X_{1i},X_{2i})}{V(X_{2i})},\\
\beta_1&=\bar{X}_{2i}-\alpha_1\cdot\bar{X}_{1i},
\label{eqA3}
\end{aligned}
\end{equation}
where
\begin{equation}
\begin{split}
C(X_{mi},X_{ni})&=\sum_{i=1}^{N}(X_{mi}-\bar{X}_{m})(X_{ni}-\bar{X}_{n}),\\
V(X_{mi})&=\sum_{i=1}^{N}(X_{mi}-\bar{X}_{m})^2,\\
\bar{X}_{m}&=\frac{1}{N}\sum_{i=1}^{N}X_{mi}.
\label{eqA4}
\end{split}
\end{equation}
Similarly, one can obtain the coefficients $\alpha_2$ and $\beta_2$
for OLS({\it X\textbar Y}), and the coefficients $\alpha_{\rm RMA}$
and $\beta_{\rm RMA}$ for OLS-RMA can be defined as \citep[for
  details, see][]{IFA+90}
\begin{equation}
\begin{split}
\alpha_{\rm RMA}&=(\alpha_1\alpha_2)^{1/2},\\
\beta_{\rm RMA}&=\bar{Y}_{2}-\alpha_{\rm RMA}\cdot\bar{Y}_{1}.
\label{eqA5}
\end{split}
\end{equation}
According to \citet{AB+96}, the fitting coefficients for the BCES
method can be obtained from the OLS method from
\begin{equation}
\begin{split}
C(Y_{mi},Y_{ni})&=C(X_{mi},X_{ni})+\sum_{i=1}^{N}\epsilon_{mi}\epsilon_{ni},\\
V(Y_{mi})&=V(X_{mi})+\sum_{i=1}^{N}\epsilon_{mi}^2,\\
\bar{Y}_{mi}&=\bar{X}_{mi}.
\label{eqA6}
\end{split}
\end{equation}
Inserting Equations (A3), (A4), and (A6) into Equations (A5), one can
obtain the fitting coefficients $\hat{\alpha}_{\rm RMA}$,
$\hat{\beta}_{\rm RMA}$ for BCES-RMA fitting.

Now we extend the method for 3D data fitting. Let the variables haveing
intrinsic real values be denoted by $X_{1i}, X_{2i}, X_{3i}$, and the
observed data by $Y_{1i}, Y_{2i}, Y_{3i}$, hence the relation between
observed data and the variables is
\begin{equation}
Y_{1i}=X_{1i}\pm \epsilon_{1i}, Y_{2i}=X_{2i}\pm \epsilon_{2i}, \rm{and~}
Y_{3i}=X_{3i}\pm \epsilon_{3i}.
\label{eqA7}
\end{equation} 
The linear regression model is formulated as
\begin{equation}
X_{3i}=\alpha' X_{1i}+\beta' X_{2i}+\gamma'.
\label{eqA8}
\end{equation}
As in 2D fitting, one can get the coefficients $\alpha_{1}'$,
$\beta_1'$, and $\gamma_1'$ for OLS($Y_3$\textbar$Y_1, Y_2$)
as follows:
\begin{equation}
\begin{aligned}
\centering
\beta_1'&=\frac{C(X_{1i},X_{3i})C(X_{1i},X_{2i})-C(X_{2i},X_{3i})V(X_{1i})}{C^2(X_{1i},X_{2i})-V(X_{1i})V(X_{2i})},\\ 
\alpha_1'&=\frac{C(X_{1i},X_{3i})-\beta_1'\cdot C(X_{1i},X_{2i})}{V_{1i}},\\ 
\gamma_1'&=\bar{X}_{3i}-\alpha_1'\cdot
\bar{X}_{1i}-\beta_1'\cdot \bar{X}_{2i}.
\label{eqA9}
\end{aligned}
\end{equation}
One can also obtain $\alpha_2'$, $\beta_2'$, and $\gamma_2'$ for
BCES($Y_{1}$\textbar$Y_{2},Y_{3}$) and $\alpha_3'$, $\beta_3'$, and
$\gamma_3'$ for BCES($Y_{2}$\textbar$Y_{1},Y_{3}$), respectively. In
principle, the 3D RMA fitting is to search for a plane that can
minimize the volume of a rectangular solid whose edges are parallel to
the axises $Y_{1}$, $Y_{2}$, and $Y_{3}$. It is not easy, however, to
obtain the fitting coefficients analytically. We define 3D OLS-RMA
fitting coefficients as
\begin{equation}
\begin{split}
\alpha_{\rm
  RMA}'&=(\alpha_{1}'\alpha_{2}'\alpha_{3}')^{1/3},\\ \beta_{\rm
  RMA}'&=(\beta_{1}'\beta_{2}'\beta_{3}')^{1/3},\\ \gamma_{\rm
  RMA}'&=\bar{Y}_{3}-\alpha_{\rm RMA}'\cdot\bar{Y}_{1}-\beta_{\rm
  RMA}'\cdot\bar{Y}_{2}.
\label{eqA10}
\end{split}
\end{equation}
Inserting Equations (A4), (A6), and (A9) into Equations (A10), one can
obtain the fitting coefficients $\hat{\alpha}_{\rm RMA}'$,
$\hat{\beta}_{\rm RMA}'$ and $\hat{\gamma}_{\rm RMA}'$. The intrinsic
scatter $\rm \sigma^2/dof$ for 3D fitting is then calculated by
\citep{CEMM14}
\begin{equation}
\frac{\sigma^{2}}{\rm dof}=\frac{\sum_{i=1}^{N}(r_{3i}-\bar{r}_{3})^2-\sum_{i=1}^{N}\epsilon_{3i}^2}{N-3},
\label{eqA11}
\end{equation}
where $r_{3i}$ is the residual
\begin{equation}
r_{3i}=Y_{3i}-\hat{\alpha}_{\rm RMA}'\cdot Y_{1i}-\hat{\beta}_{\rm
  RMA}'\cdot Y_{2i}-\gamma_{\rm RMA}',
\label{eqA12}
\end{equation}
then $\rm \chi^2/dof$ can be written as 
\begin{equation}
  \frac{\chi^2}{\rm dof}=\frac{1}{N-3}\sum_{i=1}^{N}\frac{r_{3i}^2}
       {\epsilon_{3i}^2+\hat{\alpha}_{\rm
           RMA}'^{2}\epsilon_{1i}^2+\hat{\beta}_{\rm
           RMA}'^{2}\epsilon_{2i}^2+\sigma^{2}/\rm dof}.
\label{eqA13}
\end{equation}
\end{appendix}

\bibliography{journals,cluster,ref}

\begin{thebibliography}{112}
\expandafter\ifx\csname natexlab\endcsname\relax\def\natexlab#1{#1}\fi

\bibitem[{{Akritas} \& {Bershady}(1996)}]{AB+96}
{Akritas}, M.~G., \& {Bershady}, M.~A. 1996, \apj, 470, 706

\bibitem[{{Arnaud} {et~al.}(2010){Arnaud}, {Pratt}, {Piffaretti},
  {B{\"o}hringer}, {Croston}, \& {Pointecouteau}}]{APP+10}
{Arnaud}, M., {Pratt}, G.~W., {Piffaretti}, R., {et~al.} 2010, \aap, 517, A92

\bibitem[{{Bacchi} {et~al.}(2003){Bacchi}, {Feretti}, {Giovannini}, \&
  {Govoni}}]{BFG+03}
{Bacchi}, M., {Feretti}, L., {Giovannini}, G., \& {Govoni}, F. 2003, \aap, 400,
  465

\bibitem[{{Basu}(2012)}]{Ba+12}
{Basu}, K. 2012, \mnras, 421, L112

\bibitem[{{B{\"o}hringer} {et~al.}(2000){B{\"o}hringer}, {Voges}, {Huchra},
  {McLean}, {Giacconi}, {Rosati}, {Burg}, {Mader}, {Schuecker}, {Simi{\c c}},
  {Komossa}, {Reiprich}, {Retzlaff}, \& {Tr{\"u}mper}}]{BVH+00}
{B{\"o}hringer}, H., {Voges}, W., {Huchra}, J.~P., {et~al.} 2000, \apjs, 129,
  435

\bibitem[{{B{\"o}hringer} {et~al.}(2004){B{\"o}hringer}, {Schuecker}, {Guzzo},
  {Collins}, {Voges}, {Cruddace}, {Ortiz-Gil}, {Chincarini}, {De Grandi},
  {Edge}, {MacGillivray}, {Neumann}, {Schindler}, \& {Shaver}}]{BSGC+04}
{B{\"o}hringer}, H., {Schuecker}, P., {Guzzo}, L., {et~al.} 2004, \aap, 425,
  367

\bibitem[{{B{\"o}hringer} {et~al.}(2010){B{\"o}hringer}, {Pratt}, {Arnaud},
  {Borgani}, {Croston}, {Ponman}, {Ameglio}, {Temple}, \& {Dolag}}]{BPA+10}
{B{\"o}hringer}, H., {Pratt}, G.~W., {Arnaud}, M., {et~al.} 2010, \aap, 514,
  A32

\bibitem[{{Bonafede} {et~al.}(2009{\natexlab{a}}){Bonafede}, {Giovannini},
  {Feretti}, {Govoni}, \& {Murgia}}]{BGFG+09}
{Bonafede}, A., {Giovannini}, G., {Feretti}, L., {Govoni}, F., \& {Murgia}, M.
  2009{\natexlab{a}}, \aap, 494, 429

\bibitem[{{Bonafede} {et~al.}(2014{\natexlab{a}}){Bonafede}, {Intema},
  {Br{\"u}ggen}, {Girardi}, {Nonino}, {Kantharia}, {van Weeren}, \&
  {R{\"o}ttgering}}]{BIBG+14}
{Bonafede}, A., {Intema}, H.~T., {Br{\"u}ggen}, M., {et~al.}
  2014{\natexlab{a}}, \apj, 785, 1

\bibitem[{{Bonafede} {et~al.}(2009{\natexlab{b}}){Bonafede}, {Feretti},
  {Giovannini}, {Govoni}, {Murgia}, {Taylor}, {Ebeling}, {Allen}, {Gentile}, \&
  {Pihlstr{\"o}m}}]{BFGG+09}
{Bonafede}, A., {Feretti}, L., {Giovannini}, G., {et~al.} 2009{\natexlab{b}},
  \aap, 503, 707

\bibitem[{{Bonafede} {et~al.}(2012){Bonafede}, {Br{\"u}ggen}, {van Weeren},
  {Vazza}, {Giovannini}, {Ebeling}, {Edge}, {Hoeft}, \& {Klein}}]{BBVV+12}
{Bonafede}, A., {Br{\"u}ggen}, M., {van Weeren}, R., {et~al.} 2012, \mnras,
  426, 40

\bibitem[{{Bonafede} {et~al.}(2014{\natexlab{b}}){Bonafede}, {Intema},
  {Br{\"u}ggen}, {Russell}, {Ogrean}, {Basu}, {Sommer}, {van Weeren},
  {Cassano}, {Fabian}, \& {R{\"o}ttgering}}]{BIB+14}
{Bonafede}, A., {Intema}, H.~T., {Br{\"u}ggen}, M., {et~al.}
  2014{\natexlab{b}}, \mnras, 444, L44

\bibitem[{{Brentjens}(2008)}]{Br08}
{Brentjens}, M.~A. 2008, \aap, 489, 69

\bibitem[{{Brown} {et~al.}(2011){Brown}, {Duesterhoeft}, \& {Rudnick}}]{BDR+11}
{Brown}, S., {Duesterhoeft}, J., \& {Rudnick}, L. 2011, \apjl, 727, L25

\bibitem[{{Brunetti} {et~al.}(2009){Brunetti}, {Cassano}, {Dolag}, \&
  {Setti}}]{BCD+09}
{Brunetti}, G., {Cassano}, R., {Dolag}, K., \& {Setti}, G. 2009, \aap, 507, 661

\bibitem[{{Brunetti} \& {Jones}(2014)}]{BJ14}
{Brunetti}, G., \& {Jones}, T.~W. 2014, International Journal of Modern Physics
  D, 23, 30007

\bibitem[{{Brunetti} {et~al.}(2007){Brunetti}, {Venturi}, {Dallacasa},
  {Cassano}, {Dolag}, {Giacintucci}, \& {Setti}}]{BVD+07}
{Brunetti}, G., {Venturi}, T., {Dallacasa}, D., {et~al.} 2007, \apjl, 670, L5

\bibitem[{{Brunetti} {et~al.}(2008){Brunetti}, {Giacintucci}, {Cassano},
  {Lane}, {Dallacasa}, {Venturi}, {Kassim}, {Setti}, {Cotton}, \&
  {Markevitch}}]{BGCL+08}
{Brunetti}, G., {Giacintucci}, S., {Cassano}, R., {et~al.} 2008, \nat, 455, 944

\bibitem[{{Buote} \& {Tsai}(1995)}]{BT+95}
{Buote}, D.~A., \& {Tsai}, J.~C. 1995, \apj, 452, 522

\bibitem[{{Cassano} {et~al.}(2010){Cassano}, {Ettori}, {Giacintucci},
  {Brunetti}, {Markevitch}, {Venturi}, \& {Gitti}}]{CEGB+10}
{Cassano}, R., {Ettori}, S., {Giacintucci}, S., {et~al.} 2010, \apjl, 721, L82

\bibitem[{{Cassano} {et~al.}(2008){Cassano}, {Gitti}, \& {Brunetti}}]{CGB+08}
{Cassano}, R., {Gitti}, M., \& {Brunetti}, G. 2008, \aap, 486, L31

\bibitem[{{Cassano} {et~al.}(2013){Cassano}, {Ettori}, {Brunetti},
  {Giacintucci}, {Pratt}, {Venturi}, {Kale}, {Dolag}, \& {Markevitch}}]{CEB+13}
{Cassano}, R., {Ettori}, S., {Brunetti}, G., {et~al.} 2013, \apj, 777, 141

\bibitem[{{Clarke} \& {Ensslin}(2006)}]{CE+06}
{Clarke}, T.~E., \& {Ensslin}, T.~A. 2006, \aj, 131, 2900

\bibitem[{{Colafrancesco} {et~al.}(2014){Colafrancesco}, {Emritte}, {Mhlahlo},
  \& {Marchegiani}}]{CEMM14}
{Colafrancesco}, S., {Emritte}, M.~S., {Mhlahlo}, N., \& {Marchegiani}, P.
  2014, \aap, 566, A42

\bibitem[{{Cuciti} {et~al.}(2015){Cuciti}, {Cassano}, {Brunetti}, {Dallacasa},
  {Kale}, {Ettori}, \& {Venturi}}]{CCB+15}
{Cuciti}, V., {Cassano}, R., {Brunetti}, G., {et~al.} 2015, \aap, 580, A97

\bibitem[{{Dallacasa} {et~al.}(2009){Dallacasa}, {Brunetti}, {Giacintucci},
  {Cassano}, {Venturi}, {Macario}, {Kassim}, {Lane}, \& {Setti}}]{DBG+09}
{Dallacasa}, D., {Brunetti}, G., {Giacintucci}, S., {et~al.} 2009, \apj, 699,
  1288

\bibitem[{{David} {et~al.}(1999){David}, {Forman}, \& {Jones}}]{DFJ+99}
{David}, L.~P., {Forman}, W., \& {Jones}, C. 1999, \apj, 519, 533

\bibitem[{{de Gasperin} {et~al.}(2014){de Gasperin}, {van Weeren},
  {Br{\"u}ggen}, {Vazza}, {Bonafede}, \& {Intema}}]{DVB+14}
{de Gasperin}, F., {van Weeren}, R.~J., {Br{\"u}ggen}, M., {et~al.} 2014,
  \mnras, 444, 3130

\bibitem[{{Drabent} {et~al.}(2015){Drabent}, {Hoeft}, {Pizzo}, {Bonafede}, {van
  Weeren}, \& {Klein}}]{DHP+15}
{Drabent}, A., {Hoeft}, M., {Pizzo}, R.~F., {et~al.} 2015, \aap, 575, A8

\bibitem[{{Ebeling} {et~al.}(2007){Ebeling}, {Barrett}, {Donovan}, {Ma},
  {Edge}, \& {van Speybroeck}}]{EBDM+07}
{Ebeling}, H., {Barrett}, E., {Donovan}, D., {et~al.} 2007, \apjl, 661, L33

\bibitem[{{Ebeling} {et~al.}(2000){Ebeling}, {Edge}, {Allen}, {Crawford},
  {Fabian}, \& {Huchra}}]{EEAC+00}
{Ebeling}, H., {Edge}, A.~C., {Allen}, S.~W., {et~al.} 2000, \mnras, 318, 333

\bibitem[{{Ebeling} {et~al.}(1998){Ebeling}, {Edge}, {Bohringer}, {Allen},
  {Crawford}, {Fabian}, {Voges}, \& {Huchra}}]{EEB+98}
{Ebeling}, H., {Edge}, A.~C., {Bohringer}, H., {et~al.} 1998, \mnras, 301, 881

\bibitem[{{Ebeling} {et~al.}(2010){Ebeling}, {Edge}, {Mantz}, {Barrett},
  {Henry}, {Ma}, \& {van Speybroeck}}]{EEMB+10}
{Ebeling}, H., {Edge}, A.~C., {Mantz}, A., {et~al.} 2010, \mnras, 407, 83

\bibitem[{{Ebeling} {et~al.}(2002){Ebeling}, {Mullis}, \& {Tully}}]{EMT+02}
{Ebeling}, H., {Mullis}, C.~R., \& {Tully}, R.~B. 2002, \apj, 580, 774

\bibitem[{{Ebeling} {et~al.}(1996){Ebeling}, {Voges}, {Bohringer}, {Edge},
  {Huchra}, \& {Briel}}]{EVB+96}
{Ebeling}, H., {Voges}, W., {Bohringer}, H., {et~al.} 1996, \mnras, 281, 799

\bibitem[{{Feretti} {et~al.}(2001){Feretti}, {Fusco-Femiano}, {Giovannini}, \&
  {Govoni}}]{FFGG+01}
{Feretti}, L., {Fusco-Femiano}, R., {Giovannini}, G., \& {Govoni}, F. 2001,
  \aap, 373, 106

\bibitem[{{Feretti} {et~al.}(2012){Feretti}, {Giovannini}, {Govoni}, \&
  {Murgia}}]{FGG+12}
{Feretti}, L., {Giovannini}, G., {Govoni}, F., \& {Murgia}, M. 2012, \aapr, 20,
  54

\bibitem[{{Feretti} {et~al.}(2004){Feretti}, {Orr{\`u}}, {Brunetti},
  {Giovannini}, {Kassim}, \& {Setti}}]{FOBG+04}
{Feretti}, L., {Orr{\`u}}, E., {Brunetti}, G., {et~al.} 2004, \aap, 423, 111

\bibitem[{{Feretti} {et~al.}(2005){Feretti}, {Schuecker}, {B{\"o}hringer},
  {Govoni}, \& {Giovannini}}]{FSBG+05}
{Feretti}, L., {Schuecker}, P., {B{\"o}hringer}, H., {Govoni}, F., \&
  {Giovannini}, G. 2005, \aap, 444, 157

\bibitem[{{Giacintucci} {et~al.}(2011{\natexlab{a}}){Giacintucci}, {Dallacasa},
  {Venturi}, {Brunetti}, {Cassano}, {Markevitch}, \& {Athreya}}]{GDVB+11}
{Giacintucci}, S., {Dallacasa}, D., {Venturi}, T., {et~al.} 2011{\natexlab{a}},
  \aap, 534, A57

\bibitem[{{Giacintucci} {et~al.}(2013){Giacintucci}, {Kale}, {Wik}, {Venturi},
  \& {Markevitch}}]{GKWV+13}
{Giacintucci}, S., {Kale}, R., {Wik}, D.~R., {Venturi}, T., \& {Markevitch}, M.
  2013, \apj, 766, 18

\bibitem[{{Giacintucci} {et~al.}(2011{\natexlab{b}}){Giacintucci},
  {Markevitch}, {Brunetti}, {Cassano}, \& {Venturi}}]{GMB+11}
{Giacintucci}, S., {Markevitch}, M., {Brunetti}, G., {Cassano}, R., \&
  {Venturi}, T. 2011{\natexlab{b}}, \aap, 525, L10

\bibitem[{{Giacintucci} {et~al.}(2014{\natexlab{a}}){Giacintucci},
  {Markevitch}, {Brunetti}, {ZuHone}, {Venturi}, {Mazzotta}, \&
  {Bourdin}}]{GMB+14}
{Giacintucci}, S., {Markevitch}, M., {Brunetti}, G., {et~al.}
  2014{\natexlab{a}}, \apj, 795, 73

\bibitem[{{Giacintucci} {et~al.}(2014{\natexlab{b}}){Giacintucci},
  {Markevitch}, {Venturi}, {Clarke}, {Cassano}, \& {Mazzotta}}]{GMV+14}
{Giacintucci}, S., {Markevitch}, M., {Venturi}, T., {et~al.}
  2014{\natexlab{b}}, \apj, 781, 9

\bibitem[{{Giacintucci} {et~al.}(2009{\natexlab{a}}){Giacintucci}, {Venturi},
  {Brunetti}, {Dallacasa}, {Mazzotta}, {Cassano}, {Bardelli}, \&
  {Zucca}}]{GVBD+09}
{Giacintucci}, S., {Venturi}, T., {Brunetti}, G., {et~al.} 2009{\natexlab{a}},
  \aap, 505, 45

\bibitem[{{Giacintucci} {et~al.}(2009{\natexlab{b}}){Giacintucci}, {Venturi},
  {Cassano}, {Dallacasa}, \& {Brunetti}}]{GVC+09}
{Giacintucci}, S., {Venturi}, T., {Cassano}, R., {Dallacasa}, D., \&
  {Brunetti}, G. 2009{\natexlab{b}}, \apjl, 704, L54

\bibitem[{{Giacintucci} {et~al.}(2005){Giacintucci}, {Venturi}, {Brunetti},
  {Bardelli}, {Dallacasa}, {Ettori}, {Finoguenov}, {Rao}, \& {Zucca}}]{GVBB+05}
{Giacintucci}, S., {Venturi}, T., {Brunetti}, G., {et~al.} 2005, \aap, 440, 867

\bibitem[{{Giacintucci} {et~al.}(2008){Giacintucci}, {Venturi}, {Macario},
  {Dallacasa}, {Brunetti}, {Markevitch}, {Cassano}, {Bardelli}, \&
  {Athreya}}]{GVM+08}
{Giacintucci}, S., {Venturi}, T., {Macario}, G., {et~al.} 2008, \aap, 486, 347

\bibitem[{{Giovannini} {et~al.}(2009){Giovannini}, {Bonafede}, {Feretti},
  {Govoni}, {Murgia}, {Ferrari}, \& {Monti}}]{GBF+09}
{Giovannini}, G., {Bonafede}, A., {Feretti}, L., {et~al.} 2009, \aap, 507, 1257

\bibitem[{{Giovannini} \& {Feretti}(2000)}]{GF+00}
{Giovannini}, G., \& {Feretti}, L. 2000, New Astro., 5, 335

\bibitem[{{Giovannini} {et~al.}(1991){Giovannini}, {Feretti}, \&
  {Stanghellini}}]{GFS+91}
{Giovannini}, G., {Feretti}, L., \& {Stanghellini}, C. 1991, \aap, 252, 528

\bibitem[{{Giovannini} {et~al.}(1993){Giovannini}, {Feretti}, {Venturi}, {Kim},
  \& {Kronberg}}]{GFV+93}
{Giovannini}, G., {Feretti}, L., {Venturi}, T., {Kim}, K.-T., \& {Kronberg},
  P.~P. 1993, \apj, 406, 399

\bibitem[{{Govoni} {et~al.}(2001){Govoni}, {Feretti}, {Giovannini},
  {B{\"o}hringer}, {Reiprich}, \& {Murgia}}]{GFG+01}
{Govoni}, F., {Feretti}, L., {Giovannini}, G., {et~al.} 2001, \aap, 376, 803

\bibitem[{{Govoni} {et~al.}(2005){Govoni}, {Murgia}, {Feretti}, {Giovannini},
  {Dallacasa}, \& {Taylor}}]{GMF+05}
{Govoni}, F., {Murgia}, M., {Feretti}, L., {et~al.} 2005, \aap, 430, L5

\bibitem[{{Hoeft} \& {Br{\"u}ggen}(2007)}]{HB+07}
{Hoeft}, M., \& {Br{\"u}ggen}, M. 2007, \mnras, 375, 77

\bibitem[{{Isobe} {et~al.}(1990){Isobe}, {Feigelson}, {Akritas}, \&
  {Babu}}]{IFA+90}
{Isobe}, T., {Feigelson}, E.~D., {Akritas}, M.~G., \& {Babu}, G.~J. 1990, \apj,
  364, 104

\bibitem[{{Kale} {et~al.}(2012){Kale}, {Dwarakanath}, {Bagchi}, \&
  {Paul}}]{KDB+12}
{Kale}, R., {Dwarakanath}, K.~S., {Bagchi}, J., \& {Paul}, S. 2012, \mnras,
  426, 1204

\bibitem[{{Kale} {et~al.}(2013){Kale}, {Venturi}, {Giacintucci}, {Dallacasa},
  {Cassano}, {Brunetti}, {Macario}, \& {Athreya}}]{KVG+13}
{Kale}, R., {Venturi}, T., {Giacintucci}, S., {et~al.} 2013, \aap, 557, A99

\bibitem[{{Kale} {et~al.}(2015){Kale}, {Venturi}, {Giacintucci}, {Dallacasa},
  {Cassano}, {Brunetti}, {Cuciti}, {Macario}, \& {Athreya}}]{KVG+15}
---. 2015, \aap, 579, A92

\bibitem[{{Kang} \& {Ryu}(2013)}]{KR+13}
{Kang}, H., \& {Ryu}, D. 2013, \apj, 764, 95

\bibitem[{{Kim} {et~al.}(1990){Kim}, {Kronberg}, {Dewdney}, \&
  {Landecker}}]{KKD+90}
{Kim}, K.-T., {Kronberg}, P.~P., {Dewdney}, P.~E., \& {Landecker}, T.~L. 1990,
  \apj, 355, 29

\bibitem[{{Kocevski} {et~al.}(2007){Kocevski}, {Ebeling}, {Mullis}, \&
  {Tully}}]{KEMT+07}
{Kocevski}, D.~D., {Ebeling}, H., {Mullis}, C.~R., \& {Tully}, R.~B. 2007,
  \apj, 662, 224

\bibitem[{{Liang} {et~al.}(2000){Liang}, {Hunstead}, {Birkinshaw}, \&
  {Andreani}}]{LHB+00}
{Liang}, H., {Hunstead}, R.~W., {Birkinshaw}, M., \& {Andreani}, P. 2000, \apj,
  544, 686

\bibitem[{{Lindner} {et~al.}(2014){Lindner}, {Baker}, {Hughes}, {Battaglia},
  {Gupta}, {Knowles}, {Marriage}, {Menanteau}, {Moodley}, {Reese}, \&
  {Srianand}}]{LBH+14}
{Lindner}, R.~R., {Baker}, A.~J., {Hughes}, J.~P., {et~al.} 2014, \apj, 786, 49

\bibitem[{{Macario} {et~al.}(2011){Macario}, {Markevitch}, {Giacintucci},
  {Brunetti}, {Venturi}, \& {Murray}}]{MMG+11}
{Macario}, G., {Markevitch}, M., {Giacintucci}, S., {et~al.} 2011, \apj, 728,
  82

\bibitem[{{Macario} {et~al.}(2010){Macario}, {Venturi}, {Brunetti},
  {Dallacasa}, {Giacintucci}, {Cassano}, {Bardelli}, \& {Athreya}}]{MVBD+10}
{Macario}, G., {Venturi}, T., {Brunetti}, G., {et~al.} 2010, \aap, 517, A43

\bibitem[{{Mantz} {et~al.}(2010){Mantz}, {Allen}, {Ebeling}, {Rapetti}, \&
  {Drlica-Wagner}}]{MAE+10}
{Mantz}, A., {Allen}, S.~W., {Ebeling}, H., {Rapetti}, D., \& {Drlica-Wagner},
  A. 2010, \mnras, 406, 1773

\bibitem[{{Mazzotta} \& {Giacintucci}(2008)}]{MG08}
{Mazzotta}, P., \& {Giacintucci}, S. 2008, \apjl, 675, L9

\bibitem[{{Menanteau} {et~al.}(2012){Menanteau}, {Hughes}, {Sif{\'o}n},
  {Hilton}, {Gonz{\'a}lez}, {Infante}, {Barrientos}, {Baker}, {Bond}, {Das},
  {Devlin}, {Dunkley}, {Hajian}, {Hincks}, {Kosowsky}, {Marsden}, {Marriage},
  {Moodley}, {Niemack}, {Nolta}, {Page}, {Reese}, {Sehgal}, {Sievers},
  {Spergel}, {Staggs}, \& {Wollack}}]{MHS+12}
{Menanteau}, F., {Hughes}, J.~P., {Sif{\'o}n}, C., {et~al.} 2012, \apj, 748, 7

\bibitem[{{Motl} {et~al.}(2005){Motl}, {Hallman}, {Burns}, \&
  {Norman}}]{MHB+05}
{Motl}, P.~M., {Hallman}, E.~J., {Burns}, J.~O., \& {Norman}, M.~L. 2005,
  \apjl, 623, L63

\bibitem[{{Murgia} {et~al.}(2010){Murgia}, {Govoni}, {Feretti}, \&
  {Giovannini}}]{MGFG+10}
{Murgia}, M., {Govoni}, F., {Feretti}, L., \& {Giovannini}, G. 2010, \aap, 509,
  A86

\bibitem[{{Pandey-Pommier} {et~al.}(2013){Pandey-Pommier}, {Richard}, {Combes},
  {Dwarakanath}, {Guiderdoni}, {Ferrari}, {Sirothia}, \& {Narasimha}}]{PRC+13}
{Pandey-Pommier}, M., {Richard}, J., {Combes}, F., {et~al.} 2013, \aap, 557,
  A117

\bibitem[{{Piffaretti} {et~al.}(2011){Piffaretti}, {Arnaud}, {Pratt},
  {Pointecouteau}, \& {Melin}}]{PAPP+11}
{Piffaretti}, R., {Arnaud}, M., {Pratt}, G.~W., {Pointecouteau}, E., \&
  {Melin}, J.-B. 2011, \aap, 534, A109

\bibitem[{{Pizzo} \& {de Bruyn}(2009)}]{PD09}
{Pizzo}, R.~F., \& {de Bruyn}, A.~G. 2009, \aap, 507, 639

\bibitem[{{Planck Collaboration} {et~al.}(2014{\natexlab{a}}){Planck
  Collaboration}, {Ade}, {Aghanim}, {Armitage-Caplan}, {Arnaud}, {Ashdown},
  {Atrio-Barandela}, {Aumont}, {Baccigalupi}, {Banday}, \& et~al.}]{PAA+14}
{Planck Collaboration}, {Ade}, P.~A.~R., {Aghanim}, N., {et~al.}
  2014{\natexlab{a}}, \aap, 571, A20

\bibitem[{{Planck Collaboration} {et~al.}(2014{\natexlab{b}}){Planck
  Collaboration}, {Ade}, {Aghanim}, {Armitage-Caplan}, {Arnaud}, {Ashdown},
  {Atrio-Barandela}, {Aumont}, {Aussel}, {Baccigalupi}, \& et~al.}]{PAA+14b}
---. 2014{\natexlab{b}}, \aap, 571, A29

\bibitem[{{Planck Collaboration} {et~al.}(2015){Planck Collaboration}, {Ade},
  {Aghanim}, {Arnaud}, {Ashdown}, {Aumont}, {Baccigalupi}, {Banday},
  {Barreiro}, {Barrena}, \& et~al.}]{PAA+15}
---. 2015, ArXiv e-prints

\bibitem[{{Poole} {et~al.}(2006){Poole}, {Fardal}, {Babul}, {McCarthy},
  {Quinn}, \& {Wadsley}}]{PFB+06}
{Poole}, G.~B., {Fardal}, M.~A., {Babul}, A., {et~al.} 2006, \mnras, 373, 881

\bibitem[{{Popesso} {et~al.}(2004){Popesso}, {B{\"o}hringer}, {Brinkmann},
  {Voges}, \& {York}}]{PBB+04}
{Popesso}, P., {B{\"o}hringer}, H., {Brinkmann}, J., {Voges}, W., \& {York},
  D.~G. 2004, \aap, 423, 449

\bibitem[{{Press} {et~al.}(1992){Press}, {Teukolsky}, {Vetterling}, \&
  {Flannery}}]{ptvf92}
{Press}, W.~H., {Teukolsky}, S.~A., {Vetterling}, W.~T., \& {Flannery}, B.~P.
  1992, {Numerical recipes in FORTRAN. The art of scientific computing}

\bibitem[{{Reiprich} \& {B{\"o}hringer}(2002)}]{RB+02}
{Reiprich}, T.~H., \& {B{\"o}hringer}, H. 2002, \apj, 567, 716

\bibitem[{{Riseley} {et~al.}(2015){Riseley}, {Scaife}, {Oozeer}, {Magnus}, \&
  {Wise}}]{RSO+15}
{Riseley}, C.~J., {Scaife}, A.~M.~M., {Oozeer}, N., {Magnus}, L., \& {Wise},
  M.~W. 2015, \mnras, 447, 1895

\bibitem[{{Santos} {et~al.}(2008){Santos}, {Rosati}, {Tozzi}, {B{\"o}hringer},
  {Ettori}, \& {Bignamini}}]{SRT+08}
{Santos}, J.~S., {Rosati}, P., {Tozzi}, P., {et~al.} 2008, \aap, 483, 35

\bibitem[{{Shimwell} {et~al.}(2014){Shimwell}, {Brown}, {Feain}, {Feretti},
  {Gaensler}, \& {Lage}}]{SBF+14}
{Shimwell}, T.~W., {Brown}, S., {Feain}, I.~J., {et~al.} 2014, \mnras, 440,
  2901

\bibitem[{{Shimwell} {et~al.}(2015){Shimwell}, {Markevitch}, {Brown},
  {Feretti}, {Gaensler}, {Johnston-Hollitt}, {Lage}, \& {Srinivasan}}]{SMB+15}
{Shimwell}, T.~W., {Markevitch}, M., {Brown}, S., {et~al.} 2015, \mnras, 449,
  1486

\bibitem[{{Storm} {et~al.}(2015){Storm}, {Jeltema}, \& {Rudnick}}]{SJR15}
{Storm}, E., {Jeltema}, T.~E., \& {Rudnick}, L. 2015, \mnras, 448, 2495

\bibitem[{{Trasatti} {et~al.}(2015){Trasatti}, {Akamatsu}, {Lovisari}, {Klein},
  {Bonafede}, {Br{\"u}ggen}, {Dallacasa}, \& {Clarke}}]{TAL+15}
{Trasatti}, M., {Akamatsu}, H., {Lovisari}, L., {et~al.} 2015, \aap, 575, A45

\bibitem[{{Vacca} {et~al.}(2014){Vacca}, {Feretti}, {Giovannini}, {Govoni},
  {Murgia}, {Perley}, \& {Clarke}}]{VFG+14}
{Vacca}, V., {Feretti}, L., {Giovannini}, G., {et~al.} 2014, \aap, 561, A52

\bibitem[{{Vacca} {et~al.}(2011){Vacca}, {Govoni}, {Murgia}, {Giovannini},
  {Feretti}, {Tugnoli}, {Verheijen}, \& {Taylor}}]{VGM+11}
{Vacca}, V., {Govoni}, F., {Murgia}, M., {et~al.} 2011, \aap, 535, A82

\bibitem[{{van Weeren} {et~al.}(2012{\natexlab{a}}){van Weeren}, {Bonafede},
  {Ebeling}, {Edge}, {Br{\"u}ggen}, {Giovannini}, {Hoeft}, \&
  {R{\"o}ttgering}}]{VBE+12}
{van Weeren}, R.~J., {Bonafede}, A., {Ebeling}, H., {et~al.}
  2012{\natexlab{a}}, \mnras, 425, L36

\bibitem[{{van Weeren} {et~al.}(2011{\natexlab{a}}){van Weeren}, {Br{\"u}ggen},
  {R{\"o}ttgering}, \& {Hoeft}}]{VBRH+11}
{van Weeren}, R.~J., {Br{\"u}ggen}, M., {R{\"o}ttgering}, H.~J.~A., \& {Hoeft},
  M. 2011{\natexlab{a}}, \mnras, 418, 230

\bibitem[{{van Weeren} {et~al.}(2011{\natexlab{b}}){van Weeren}, {Br{\"u}ggen},
  {R{\"o}ttgering}, {Hoeft}, {Nuza}, \& {Intema}}]{VBR+11}
{van Weeren}, R.~J., {Br{\"u}ggen}, M., {R{\"o}ttgering}, H.~J.~A., {et~al.}
  2011{\natexlab{b}}, \aap, 533, A35

\bibitem[{{van Weeren} {et~al.}(2011{\natexlab{c}}){van Weeren}, {Hoeft},
  {R{\"o}ttgering}, {Br{\"u}ggen}, {Intema}, \& {van Velzen}}]{VHR+11}
{van Weeren}, R.~J., {Hoeft}, M., {R{\"o}ttgering}, H.~J.~A., {et~al.}
  2011{\natexlab{c}}, \aap, 528, A38

\bibitem[{{van Weeren} {et~al.}(2010){van Weeren}, {R{\"o}ttgering},
  {Br{\"u}ggen}, \& {Hoeft}}]{VRB+10}
{van Weeren}, R.~J., {R{\"o}ttgering}, H.~J.~A., {Br{\"u}ggen}, M., \& {Hoeft},
  M. 2010, Science, 330, 347

\bibitem[{{van Weeren} {et~al.}(2012{\natexlab{b}}){van Weeren},
  {R{\"o}ttgering}, {Intema}, {Rudnick}, {Br{\"u}ggen}, {Hoeft}, \&
  {Oonk}}]{VRI+12}
{van Weeren}, R.~J., {R{\"o}ttgering}, H.~J.~A., {Intema}, H.~T., {et~al.}
  2012{\natexlab{b}}, \aap, 546, A124

\bibitem[{{van Weeren} {et~al.}(2009){van Weeren}, {R{\"o}ttgering}, {Bagchi},
  {Raychaudhury}, {Intema}, {Miniati}, {En{\ss}lin}, {Markevitch}, \&
  {Erben}}]{VRB+09}
{van Weeren}, R.~J., {R{\"o}ttgering}, H.~J.~A., {Bagchi}, J., {et~al.} 2009,
  \aap, 506, 1083

\bibitem[{{van Weeren} {et~al.}(2014){van Weeren}, {Intema}, {Lal},
  {Andrade-Santos}, {Br{\"u}ggen}, {de Gasperin}, {Forman}, {Hoeft}, {Jones},
  {Nuza}, {R{\"o}ttgering}, \& {Stroe}}]{VILA+14}
{van Weeren}, R.~J., {Intema}, H.~T., {Lal}, D.~V., {et~al.} 2014, \apjl, 786,
  L17

\bibitem[{{Venturi} {et~al.}(2003){Venturi}, {Bardelli}, {Dallacasa},
  {Brunetti}, {Giacintucci}, {Hunstead}, \& {Morganti}}]{VBDB+03}
{Venturi}, T., {Bardelli}, S., {Dallacasa}, D., {et~al.} 2003, \aap, 402, 913

\bibitem[{{Venturi} {et~al.}(2007){Venturi}, {Giacintucci}, {Brunetti},
  {Cassano}, {Bardelli}, {Dallacasa}, \& {Setti}}]{VGBC+07}
{Venturi}, T., {Giacintucci}, S., {Brunetti}, G., {et~al.} 2007, \aap, 463, 937

\bibitem[{{Venturi} {et~al.}(2008){Venturi}, {Giacintucci}, {Dallacasa},
  {Cassano}, {Brunetti}, {Bardelli}, \& {Setti}}]{VGDC+08}
{Venturi}, T., {Giacintucci}, S., {Dallacasa}, D., {et~al.} 2008, \aap, 484,
  327

\bibitem[{{Venturi} {et~al.}(2013){Venturi}, {Giacintucci}, {Dallacasa},
  {Cassano}, {Brunetti}, {Macario}, \& {Athreya}}]{VGDC+13}
---. 2013, \aap, 551, A24

\bibitem[{{Venturi} {et~al.}(1990){Venturi}, {Giovannini}, \&
  {Feretti}}]{VGF+90}
{Venturi}, T., {Giovannini}, G., \& {Feretti}, L. 1990, \aj, 99, 1381

\bibitem[{{Vikhlinin} {et~al.}(2009){Vikhlinin}, {Burenin}, {Ebeling},
  {Forman}, {Hornstrup}, {Jones}, {Kravtsov}, {Murray}, {Nagai}, {Quintana}, \&
  {Voevodkin}}]{VBE+09}
{Vikhlinin}, A., {Burenin}, R.~A., {Ebeling}, H., {et~al.} 2009, \apj, 692,
  1033

\bibitem[{{Voges} {et~al.}(1999){Voges}, {Aschenbach}, {Boller},
  {Br{\"a}uninger}, {Briel}, {Burkert}, {Dennerl}, {Englhauser}, {Gruber},
  {Haberl}, {Hartner}, {Hasinger}, {K{\"u}rster}, {Pfeffermann}, {Pietsch},
  {Predehl}, {Rosso}, {Schmitt}, {Tr{\"u}mper}, \& {Zimmermann}}]{VAB+99}
{Voges}, W., {Aschenbach}, B., {Boller}, T., {et~al.} 1999, \aap, 349, 389

\bibitem[{{Wei{\ss}mann} {et~al.}(2013){Wei{\ss}mann}, {B{\"o}hringer}, {{\v
  S}uhada}, \& {Ameglio}}]{WBSA+13}
{Wei{\ss}mann}, A., {B{\"o}hringer}, H., {{\v S}uhada}, R., \& {Ameglio}, S.
  2013, \aap, 549, A19

\bibitem[{{Wen} \& {Han}(2013)}]{WH+13}
{Wen}, Z.~L., \& {Han}, J.~L. 2013, \mnras, 436, 275

\bibitem[{{Wen} \& {Han}(2015)}]{WH15}
---. 2015, \apj, 807, 178

\bibitem[{{Wen} {et~al.}(2012){Wen}, {Han}, \& {Liu}}]{WHL+12}
{Wen}, Z.~L., {Han}, J.~L., \& {Liu}, F.~S. 2012, \apjs, 199, 34

\bibitem[{{Wik} {et~al.}(2008){Wik}, {Sarazin}, {Ricker}, \&
  {Randall}}]{WSR+08}
{Wik}, D.~R., {Sarazin}, C.~L., {Ricker}, P.~M., \& {Randall}, S.~W. 2008,
  \apj, 680, 17

\bibitem[{{Zhao} {et~al.}(2013){Zhao}, {Jia}, {Chen}, {Li}, {Song}, \&
  {Xie}}]{ZJC+13}
{Zhao}, H.-H., {Jia}, S.-M., {Chen}, Y., {et~al.} 2013, \apj, 778, 124

\bibitem[{{Zhao} {et~al.}(2015){Zhao}, {Li}, {Chen}, {Jia}, \& {Song}}]{ZLC+15}
{Zhao}, H.-H., {Li}, C.-K., {Chen}, Y., {Jia}, S.-M., \& {Song}, L.-M. 2015,
  \apj, 799, 47

\bibitem[{{ZuHone} {et~al.}(2013){ZuHone}, {Markevitch}, {Brunetti}, \&
  {Giacintucci}}]{ZMB+13}
{ZuHone}, J.~A., {Markevitch}, M., {Brunetti}, G., \& {Giacintucci}, S. 2013,
  \apj, 762, 78

\end{thebibliography}

\end{document}